\documentclass[amsmath,amssymb,twocolumn,10pt]{revtex4-1}
\usepackage[table]{xcolor}
\usepackage{tabularx}
\usepackage{amsmath}
\usepackage{bm}
\usepackage{graphicx}
\usepackage{hyphenat}

\usepackage[english]{babel}
\usepackage{amssymb}
\usepackage{float} 
\usepackage{subfigure} 
\usepackage{setspace} 
\usepackage{fixmath} 

\usepackage{color}
\definecolor{darkblue}{rgb}{0,0,0.5}
\definecolor{lila}{rgb}{0.3,0,0.3}
\definecolor{turq}{rgb}{0,0.1,0.4}
\definecolor{lightblue}{rgb}{0.7,0.7,0.9}
\usepackage{url} 
\usepackage[pdftex,
 colorlinks=true,
 backref=page,
 linkcolor=darkblue, 
 filecolor=red,
 citecolor=turq, 
 urlcolor=lila, 
 pdftitle={Better Randomness with Single Photons},
 pdfauthor={Lukas Oberreiter and Ilja Gerhardt},
 pdfsubject={},
 pdfkeywords={}
 pdfpagelabels=true, 
 breaklinks=false,
 plainpages=false,
 backref=false,
 bookmarks,
 bookmarksnumbered=true]{hyperref}

\newcommand{\adet}{\mathbf{a}}
\newcommand{\bdet}{\mathbf{b}}

\makeatletter
\DeclareRobustCommand\citenum
   {\begingroup
     \NAT@swatrue\let\NAT@ctype\z@\NAT@parfalse\let\textsuperscript\relax
     \NAT@citexnum[][]}
\makeatother

\begin{document}

\title{\Large Better Randomness with Single Photons}

\author{Lukas Oberreiter}

\affiliation{3. Physikalisches Institut, Universit\"at Stuttgart and Stuttgart Research Center of Photonic Engineering (SCoPE), Pfaffenwaldring 57, Stuttgart, D-70569, Germany}

\author{Ilja Gerhardt}
\email{i.gerhardt@fkf.mpg.de}
\affiliation{3. Physikalisches Institut, Universit\"at Stuttgart and Stuttgart Research Center of Photonic Engineering (SCoPE), Pfaffenwaldring 57, Stuttgart, D-70569, Germany}
\affiliation{Max Planck Institute for Solid State Research, Heisenbergstra\ss e 1, D-70569 Stuttgart, Germany}

\begin{abstract}
Randomness is one of the most important resources in modern information science, since encryption founds upon the trust in random numbers~\cite{lenstra__2012}. Since it is impossible to prove if an existing random bit string is truly random, it is relevant that they be generated in a trust worthy process. This requires specialized hardware for random numbers, for example a die~\cite{galton_n_1890} or a tossed coin~\cite{feller__1968}. But when all input parameters are known, their outcome might still be predicted~\cite{laplace__1951}. A quantum mechanical superposition allows for provably true random bit generation~\cite{schmidt_joap_1970,erber_n_1985}. In the past decade many quantum random number generators (QRNGs) were realized~\cite{stefanov_jmo_2000,jennewein_rosi_2000,gabriel_np_2010,abellan_oe_2014,sanguinetti_prx_2014}. A photonic implementation is described as a photon which impinges on a beam splitter~\cite{frauchiger_a_2013}, but such a protocol is rarely realized with non-classical light~\cite{bronner_ejop_2009,branning_josab_2010,graefe_nphot_2014} or anti-bunched single photons. Instead, laser sources or light emitting diodes~\cite{stefanov_a_1999,jennewein_rosi_2000,quantique_w_2010} are used. Here we analyze the difference in generating a true random bit string with a laser and with anti-bunched light. We show that a single photon source provides more randomness than even a brighter laser. This gain of usable entropy proves the advantages of true single photons versus coherent input states of light in an experimental implementation. The underlying advantage can be adapted to microscopy and sensing.
\end{abstract}

\maketitle

\begin{figure*}[t!]
  \includegraphics[width=\textwidth]{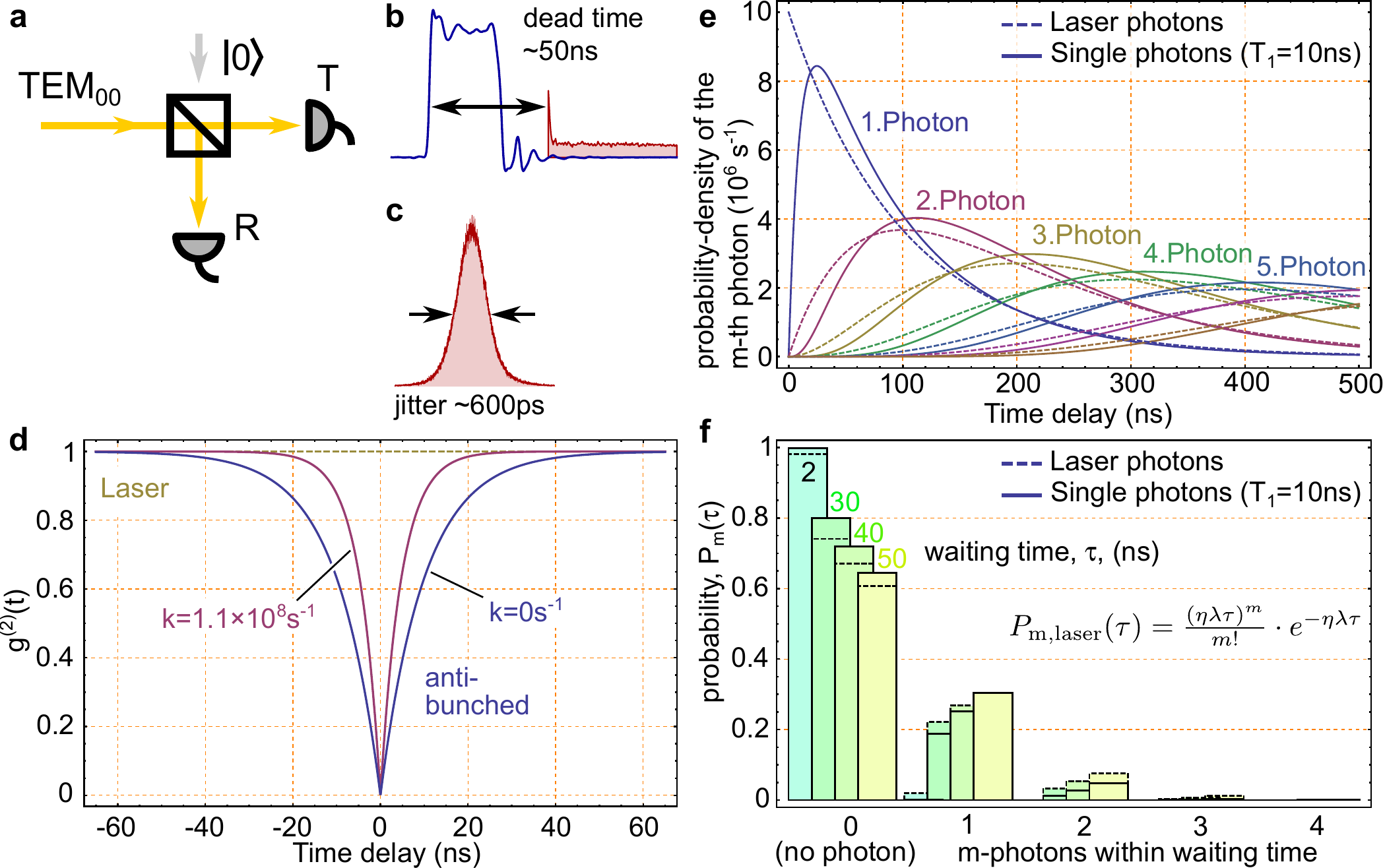}
  \caption{\textbf{a)} Incoming photons impinge onto a beam splitter. The output is equipped with single photon detectors. The outputs are labeled as T (transmitted) and R (reflected) \textbf{b)} After a detection event, the probability of detection is reduced to zero for some time delay. A histogram (red) of the next detection event after a pulse. The detector can detect the next photon after 50~ns. \textbf{c)} Timing jitter of the setup, when exposed to a pulsed laser. This is the convolution of both detector jitters. \textbf{d)} Photon statistics of anti-bunched light, emitted from a single emitter and a laser. The single emitter is displayed at different excitation rates ($k$). $\mathit{\Gamma}$ is assumed to be 1/(10~ns). \textbf{e)} Temporal probability density of the $m$th photon, after the emission of a single photon, $L_{\mathrm{m}}(\tau)$. \textbf{f)} Probability distribution, $P_{\mathrm{m}}(\tau)$, of certain photon numbers within a pre-defined waiting time. Conditioned on the emission of a single photon at time $t=0$. The flux of photons, $\lambda$, was assumed to be 10$^7$ photons per second. No attenuation was assumed ($\eta = 1$).}
  \label{fig:setup}
\end{figure*}

A basic photonic quantum random bit generator is described as follows: a (single) photon imposes on a beam splitter and is either reflected ($R$) or transmitted ($T$) (Fig.~\ref{fig:setup}a). Behind every output port a single photon detector is placed, generating an electrical pulse for an incoming photon. The generator has two outcomes $\adet,\bdet \in \{R,T\}$ and $\bdet \neq \adet$. Experimentally, the incoming light commonly does not consist of single photons, but a coherent state of a laser. This can be described as several copies of the same pure state~\cite{enk_prl_2001}, sufficient to realize a quantum superposition.

The difference between a laser and a single photon source on a beam splitter is denoted as $|0\rangle_1 |\alpha \rangle_2 \rightarrow 1/\sqrt{2} |i \alpha\rangle_{\mathrm{R}} |\alpha\rangle_{\mathrm{T}}$ for a laser, with the average photon number $| \alpha |^2 = \langle n\rangle = \lambda$. The single photon source is described by $|0\rangle_1|1\rangle_2 \rightarrow 1/\sqrt{2} \left[i |1\rangle_{\mathrm{R}}|0\rangle_{\mathrm{T}} + |0\rangle_{\mathrm{R}}1\rangle_{\mathrm{T}}\right]$. Furthermore, the temporal photon statistics differs. A single photon source obeys an ``anti-bunched'' photon statistics, proving that no two photons are emitted at the same time (Fig.~\ref{fig:setup}d). The coherent state of a laser shows a $g^{(2)}(\tau)$ function of unity for all delay times $\tau$. Therefore, the arrival times of single photons on the randomness device are different. Here we compare the different outcomes of the QRNG with different photonic input states. We disregard all other sources of side-information known to an external adversary.

The normalized single photon correlation function reads as 

\begin{eqnarray}
g^{(2)}(\tau)&=&\frac{\langle I(t) I(t+\tau)\rangle}{\langle I(t)\rangle^2}\\
&=&\frac{\rho_{22}(\tau)}{\lim_{\tau \rightarrow \infty} \rho_{22}(\tau)}=1-e^{-(k+\mathit{\Gamma})|\tau|} \ .
\end{eqnarray}

\noindent
$k$ denotes the pumping rate, $\mathit{\Gamma}=1/T_1$ the inverse of the longitudinal lifetime of the emitter, and $\rho_{22}$ the excited state population. The emission rate $\lambda$ is given through the probability, that the system will be in the exited state at thermal equilibrium multiplied with $\mathit{\Gamma}$:

\begin{eqnarray}
\lambda(k) &=& \mathit{\Gamma} \cdot \lim_{\tau \rightarrow \infty} \rho_{22}(\tau)=\frac{k\mathit{\Gamma}}{k+\mathit{\Gamma}} \ .
\label{eq:lambda}
\end{eqnarray}

\noindent
This shows the typical saturation behavior of a two level system. At high count-rates the so-called ``anti bunching dip'' is narrow. At $k \rightarrow \infty$ the resulting photon statistics of a laser and a single photon source are equal. This is evident in the following plots, where the laser and single photon source emission forms one point for a emission rate of $1/T_{\mathrm{1}}=\mathit{\Gamma}$ photons per second. We assume a $T_{\mathrm{1}}$-time of 10~ns, a value corresponding to single organic dye molecules, colloidal nano crystals, and the negatively charged nitrogen-vacancy center. The maximum emission rate is 10$^8$ photons per second. This capped brightness seems to be disadvantageous for the generation of random bits, since a laser can be assumed to generate more generator outcomes per unit time, due to its higher brightness. The (anti-)correlation of subsequent detection events of a single emitter might also introduce some correlations, such that one detection event might introduce some predictability for the next detector outcome.

A single photon detector introduces many technical implications before a photon is converted into an electrical ``click''. First, single photon detection is limited by a detector quantum efficiency $\eta_{\mathrm{qe}}$. When avalanche photo diodes (APDs, e.g.\ `Excellitas') are used, the quantum efficiency is above 60\% in a spectral range of 550 to 800~nm~[\citenum{optoelectronics__2001}]. The finite temporal length of an electrical click and an active quenching operation delay the next detection event (Fig.~\ref{fig:setup}b). This is the ``dead time'' of the photon detector, $\tau_{\mathrm{dead}}$. The conversion process introduces some time delay and, more importantly, electrical jitter. This is a fluctuating delay between an incoming photon and a generated electrical pulse. The temporal spread is described by $\sigma_{\mathrm{jitt}}$ (Fig.~\ref{fig:setup}c). The histogram of joint detection events is a Gaussian distribution with a long tail~\cite{tan_tajl_2014}. This is the convolution of both jitter functions of each of the detectors alone, and measured to be 250~ps. There are more subtleties with single photon detectors, such as dark counts, after-pulsing and a break-down flash. These effects are not considered in the following, since they do not become relevant to the description here.

Electrical jitter will lead to the fact that the subsequent order of some events might be confused. It is not clear, if the event $\adet$ and then the event $\bdet$ was detected, or if it was vice versa. To exclude this ambiguity, we introduce a \emph{coincidence window}, $\tau_{\mathrm{cw}}$, which is started with every succesful detection event. A coincidence is observed, when the other detector clicks within $\tau_{\mathrm{cw}}$. To determine an upper bound of the permutation probability, $\epsilon$, we calculate the one-sided bound, that two simultaneous incoming photons are detected as subsequent events:

\begin{equation}
\epsilon=\int_{\tau=-\infty}^{-\tau_{\mathrm{cw}}}\frac{1}{\sigma_{\mathrm{jitt}} \sqrt{2 \pi}}e^{-\frac{1}{2}\left(\frac{\tau}{\sigma_{\mathrm{jitt}}}\right)^2} \, \mathrm{d} \tau =\frac{1}{2}\mbox{erfc}\left(\frac{\tau_{\mathrm{cw}}}{\sqrt{2}\sigma_{\mathrm{jitt}}}\right) \ .
\end{equation}

\noindent
We choose a $\tau_{\mathrm{cw}}$ of 2~ns. This is chosen such that the probability $\epsilon$ to confuse events is lower than 1 confusion per year at the maximal possible click rate (1/$\tau_{\mathrm{dead}}$). We could discard these events, but since they contain some usable entropy, their inclusion will lead to more extractable random bits. 

For comparison, we consider the temporal photon detection probability in the experimental configuration. With a different $g^{(2)}$-function the waiting time distribution is changed. Commonly, the photon statistics is measured in a start-stop fashion, i.e. the difference in arrival times of two subsequent events on the detectors are histogrammed. This does not result in the the real $g^{(2)}(\tau)$-function. The latter would be a histogram over \emph{all} following photon arrival times. The transition between both functions is introduced by the $JK$-formalism~\cite{reynaud_a_1983}: $J(\tau):=g^{(2)}(\tau) \cdot \eta \lambda$ defines the number density for the emission of \emph{any} following photon at time $\tau$\footnote{$\eta$ is still defined as unity when only the emission properties are analyzed, but will become relevant later}. Here, $\eta$ accounts for any losses and efficiencies in the system. $K(\tau)$ describes the probability density for the \emph{next} photon.

Mathematically they are related by~\cite{reynaud_a_1983}:

\begin{equation}
\quad K(\tau)=\mathcal{L}^{-1}[\frac{\mathcal{L}[J(\tau)]}{1+\mathcal{L}[J(\tau)]}] \ ,
\end{equation}

\noindent
where $\mathcal{L}[f]$ denotes the Laplace transform of the function $f$. With the function $K(\tau)$ it is possible to derive in an iterative way to the waiting time distribution for the $m$th photon ($L_{m}(\tau)$). For a laser it is straight forward to find an analytical expression. For the single photon source, the calculation for $L_{m}(\tau)$ is given in the supplementary material. The probability density, that the $(m+1)$th photon will be emitted at time $\tau$:

\begin{equation}
L_{m+1}(\tau)=\int_{t=0}^{\tau}L_{m}(\tau-t)K(t)\ \mathrm{d}t \quad \mbox{with} \quad L_{1}(\tau)=K(\tau)
\end{equation}

\noindent
The probability $P_m(\tau)$ to emit $m$ photons within a waiting time $\tau$ right after the emission of a photon is calculated as:

\begin{eqnarray}
P_0(\tau)&=&1-\int_{t=0}^\tau K(t)\,\mathrm{d}t \\
P_m(\tau)&=&\int_{t=0}^\tau L_m(t) P_0(\tau-t)\,\mathrm{d}t\ .
\end{eqnarray}

\noindent
When the $P_m(\tau)$ is calculated for laser light, it represents the Poisson distribution (Eqn. in Fig.~\ref{fig:setup}f). The difference between $J$ and $K$ is rarely discussed in the study of single photon emitters, where often start-stop type experiments are performed when photon anti-bunching is recorded. Only few papers include the relevant math~\cite{fleury_prl_2000,fleury_jol_2001}. When a low collection and detection probability is assumed, the $g^{(2)}(\tau)$-function is approached with a start-stop measurement. Fig.~\ref{fig:setup}e shows the waiting time distribution for the $m$-th photon. The emission of the next laser photon is a Bernoulli trial and shows an exponential behavior. The emission probability of a subsequent photon does not depend on a prior photon emission. On the other hand, a true single photon emitter will have some waiting time until it will reach the same probability. After this, it exceeds the photon emission probability of a laser.

\begin{figure}[bh]
  \includegraphics[width=\columnwidth]{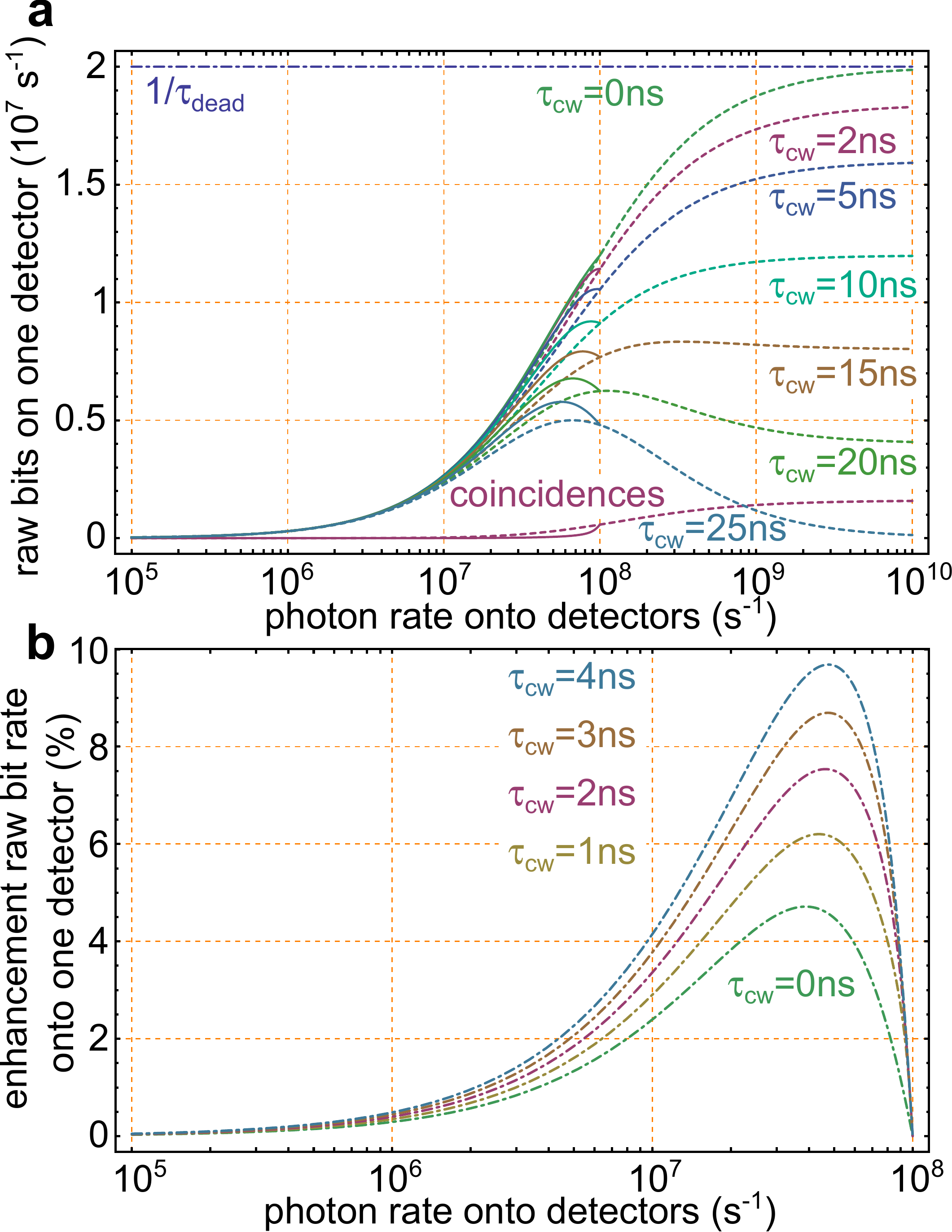}
  \caption{\textbf{a)} Raw bit rate, $\lambda_{\mathrm{bit}}^{\adet}$, onto a single detector, when the incident light onto the setup is varied. The maximum detection rate is limited to $1/\tau_{\mathrm{dead}}$. When coincidences are discarded, the number of outcomes is reduced at high photon rates. The coincidence window is varied. The amount of coincidences for $\tau_{\mathrm{cw}}=\,$2~ns is displayed at the bottom. \textbf{b)} Ratio of produced raw bit rate of a single photon source against the rate with a laser source. Coincidences are discarded, since the ambiguity of confusing them with subsequent clicks on T and R. For $\tau_{\mathrm{cw}}=\,$0~ns no coincidences occur. Still, the single photon source produces more clicks, due to the photon waiting time distribution and the detector intrinsic dead time.}
  \label{fig:rawbit}
\end{figure}

Fig.~\ref{fig:setup}f shows the emission probability within a defined waiting time. By definition a single photon source has a higher probability to emit no second photon within a certain time. The dead-time of the photon detector is in the same order of magnitude ($\tau_{\mathrm{dead}}$=50~ns). Therefore, it is likely that for a defined brightness, $\lambda$, a single photon source might produce more detection events.

The above only considers the photon \emph{emission} properties ($\eta$=1). We now introduce the technical details on the \emph{detection} of the supplied photons. We introduce the transition $\eta \rightarrow \eta_\adet:=\eta_{\mathrm{qe}} p_\adet$, where $\eta_{\mathrm{qe}}$ denotes the quantum efficiency and $p_\adet$ the probability that the photon will go from the beam splitter to detector $\adet$ and assume a ``fair'' beam splitter with a 50:50 splitting ratio. With this correction, the number of incoming photons is transferred to the number of click-producing photons when the detector is not ``dead''. The click-rate $\lambda_{\mathrm{click}}^{\adet}$ of detector $\adet$ has to be calculated. The average number of photons which will arrive within a dead-time on detector $\adet$ and \emph{would} produce a click is

\begin{equation}
m^\adet=\int_0^{\tau_{\mathrm{dead}}} J^{\adet}(\tau) \ d\tau \ .
\label{eq:m_adet}
\end{equation}

\noindent
One click on detector $\adet$ leads to one dead-time. That means that from $\lambda^\adet:=\lambda \eta_\adet$ possibly seen incoming photons to detector $\adet$, $\lambda_{\mathrm{click}}^{\adet}$ photons will produce a click and $\lambda_{\mathrm{click}}^{\adet} \cdot m^\adet$ will enter while the detector is blind. The click-rate $\lambda_{\mathrm{click}}^{\adet}$ is

\begin{equation}
\lambda^\adet= \lambda_{\mathrm{click}}^{\adet} (1+m^\adet) \qquad \Leftrightarrow \qquad \lambda_{\mathrm{click}}^{\adet}=\frac{\lambda^{\adet}}{1+m^\adet} \ .
\label{eq:lambda_2}
\end{equation}

\noindent
The outcome of a click in detector $\adet$ at time $\tau$=0 has two possible outcomes: it is the only click within the time interval $[-\tau_{\mathrm{cw}},\tau_{\mathrm{cw}}]$, or detector $\bdet$ clicks too and a coincidence event ($\adet\bdet$) is observed. The probability of detector $\bdet$ to click within $[-\tau_{\mathrm{cw}},\tau_{\mathrm{cw}}]$ is well approximated with:

\begin{equation}
P_{\mathrm{coinc}}^\bdet=\lambda_{\mathrm{click}}^\bdet \int_{-\tau_{\mathrm{cw}}}^{\tau_{\mathrm{cw}}} g^{(2)}(\tau) \,\mathrm{d}\tau \ .
\end{equation}

\noindent
This leads to a coincidence rate of $\lambda_{\mathrm{coinc}} =\lambda_{\mathrm{click}}^\adet P^\bdet_{\mathrm{coinc}}$~\footnote{Please note, that this definition deviates by a factor of two from many references, where a coincidence rate of two random sources is given as $r_{\mathrm{coinc}}=r_1 r_2 \tau_{\mathrm{cw}}$. In our case a $\tau_{\mathrm{cw}}$ of 2~ns implies, that two subsequent events $\adet,\bdet$ will be regarded as a coincidence, if their time distance is smaller than 2~ns. This makes the underlying calculation of the combinatoric conditional probabilities more intuitive. The observation of two subsequent clicks with a distance of e.g. 3~ns would not lead to a coincicence. Therefore, our $\tau_{\mathrm{cw}}$ defines a ``observation window'', which is started with the successful detection of a photon.} and a raw bit rate of detector $\adet$ defined as

\begin{equation}
\lambda_{\mathrm{bit}}^\adet = \lambda_{\mathrm{click}}^\adet \left( 1- P^\bdet_{\mathrm{coinc}} \right) \ .
\end{equation}

\noindent
When discarding coincidences, the total raw bit rate $\lambda_{\mathrm{bit}}$ is the sum over both detectors:

\begin{equation}
\lambda_{\mathrm{bit}}
=\lambda_{\mathrm{bit}}^\adet+\lambda_{\mathrm{bit}}^\bdet
=\lambda_{\mathrm{click}}^\adet+\lambda_{\mathrm{click}}^\bdet-2\lambda_{\mathrm{coinc}} \ .
\end{equation}

\noindent
Fig.~\ref{fig:rawbit}a shows the bit rates on detector $\adet$. Depending on the photon rate, the dead time influences the probability of detection. For high photon rates, the coincidence rate increases and levels off at the ratio of $\tau_{\mathrm{dead}} / (2 \tau_{\mathrm{cw}})$. For infinite operation of the QRNG, the resulting ratio of single events to coincidences will be a fixed ratio.

\begin{figure*}[bh]
  \includegraphics[width=\textwidth]{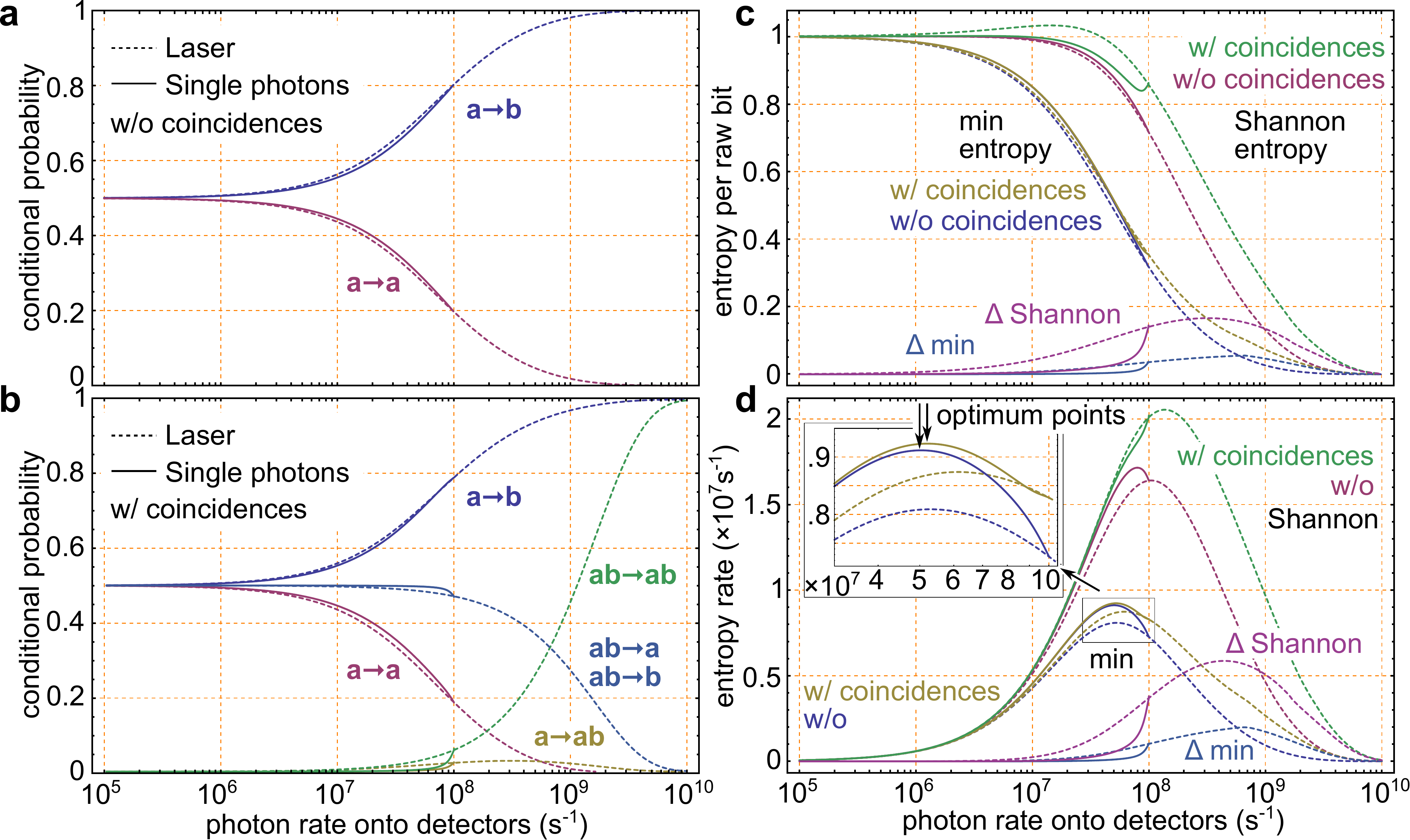}
  \caption{To calculate the final entropy, all conditional probabilities of subsequent raw outcomes have to be determined. Everything is conditioned to the fact that detector $\adet$ produced a click. This might be a single click ($\adet$), or a coincidence ($\adet\bdet$). The next detection event can be the same detector ($\adet$), the other detector ($\bdet$) or a coincidence ($\adet \bdet$). When coincidence events are discarded (\textbf{a)}), only the curves $\adet \rightarrow \bdet$ and $\adet \rightarrow \adet$ remain and are symmetric around 0.5. \textbf{b)} Conditional probabilities, including the coincidences. All probabilities for one start condition add up to unity. \textbf{c)} Calculated conditional Shannon- and min-entropy per detection event. Both cases are included, with and without coincidences as output events. Coincidences give an additional source of entropy, and result in ``excess'' bits, resulting in a Shannon entropy above unity. The difference of the entropy is shown in the bottom curves, labeled as $\Delta$. A single photon source has more entropy, when the min-entropy is calculated. \textbf{d)} The final entropy rate (outcome rate $\times$ entropy per outcome). Close to saturation at 10$^8$ incident photons onto the device, the min-entropy peaks. This is the optimum working condition of the random bit generator (zoomed inset). The single photon source results in 12.6\% (without coincidences) or 5.6\% (with coincidences) more extractable random bits than the laser under optimum working conditions.}
  \label{fig:entropy}
\end{figure*}

Fig.~\ref{fig:rawbit}b shows the increased bit rate between a single photon source and a laser. At high photon flux, the detector spends more time in the ``dead'' state, and therefore the bit rate is higher for the single photon source. The flux density is reduced within the dead time of the detector and only later increased. This effect gets more relevant, when coincidence events are discarded to avoid a bit permutation.

For any randomness generation, the relevant information is the entropy content of the fundamental process of photon detection. In the case of a randomness generator the conditional Shannon entropy, accounting for the correlation of subsequent raw bits, is defined as:

\begin{eqnarray}
\mathcal{H}_\text{Sh}(X|Y)&=&\sum_y p(y) \mathcal{H}_\text{Sh}(X|Y=y)\\
&=&-\sum_y p(y) \sum_x p(x|y) \log_2{p(x|y)} \ .
\end{eqnarray}

\noindent
With two outcomes, the maximal entropy is unity. This defines an optimistic upper bound for the entropy fraction of the generator. More genralized, the conditional R\'enyi entropy has to be considered. Its lower bound is the conditional min-entropy. This gives the conservative bound for the entropy fraction in the system. It is defined as

\begin{equation}
\mathcal{H}_\text{min}(X|Y) =  -\log_2 \left[ \sum_y p(y) \max_x \left\lbrace p(x|y)  \right\rbrace \right] \ .
\end{equation}

\noindent
$p(y)$ is the probability that an arbitrary outcome has the value $y$. It is worthwhile noting that we changed the notation from $\adet$ and $\bdet$ to $x$ and $y$, since we allow for coincidences. Therefore $x \in {\adet,\bdet,\adet\bdet}$. $p(x|y)$ is the conditional probability that after the outcome $y$ the following outcome takes the value $x$.

When coincidences are discarded, the event $\adet$ is assumed first, and then the subsequent outcome $x \in \adet,\bdet$ (Fig.~\ref{fig:entropy}a). At high photon rates the transition $\adet \rightarrow \bdet$ is more likely, due to the increased probability of receiving a subsequent photon within the dead time of detector $\adet$. At very high incident photon flux, the detectors click alternately, since at each time a detector is `alive' it receives the next photon. The resulting time-ordered outcome would be $\adet,\bdet,\adet,\bdet,\adet,\bdet,\adet\ldots$. With two possible outcomes, the resulting conditional probabilities are symmetric around 0.5.

The calculation of the conditional probabilities is a combinatoric problem. For example, the transition of $\adet$ to $\bdet$ can occur, when $\adet$ clicked, and $\bdet$ occurs not within the coincidence window. Then there are several options how $\bdet$ can occur, such as within and also outside the dead-time of detector $\adet$. Equivalently, for the case $p(\adet|\adet)$ it is required that after $\adet$, detector $\bdet$ does not click before $\adet$ clicks again. A full derivation of all conditional probabilities is given in the supplementary material. In the following we extend these assumptions to the case including the coincidences.

Fig.~\ref{fig:entropy}b shows the calculation of the transition probabilities for the case, that coincidences are not discarded. When a click on $\adet$ is detected, a third outcome, namely to a coincidence ($\adet\bdet$), is possible. For simplicity, we assume that the dead time after a coincidence event ends exactly at the same time. The coincidence probability becomes relevant at high photon rates. All three conditional probabilities add up to one, but are not symmetric anymore. For very high photon rates, the probability to result in a coincidence again is reduced, although the number of coincidences is still increasing as shown in Fig.~\ref{fig:rawbit}a. This becomes clear, when we consider the conditions for a click $\adet$: It implies, that the other detector is likely in its ``dead time'', and will likely receive a next photon, when detector $\adet$ is still ``dead''.

Additionally, the outcomes for starting the calculation with a coincidence ($\adet \bdet$) are depicted. The transition probabilities from a coincidence to one of the outcomes ($\adet$ or $\bdet$) is the same. This transition becomes less likely, supported with the same arguments as for the transition of an alternating sequence of $\adet,\bdet,\adet,\bdet,\adet,\bdet,\adet \ldots$. It is important to realize that although Fig.~\ref{fig:rawbit}a shows a fixed ratio of coincidences to single clicks, the \emph{transition} from one to the other state becomes less likely at high photon rates.

With all transition probabilities, we calculate the entropy per raw bit. When few photons are coming, their temporal spacing is large against the dead time or the influence of the photon statistics. Also, the number of coincidences is negligible. Therefore, at low photon rates, the entropy per incoming photon is close to unity, and the generator will pass all statistical randomness tests also without any post-processing~\cite{graefe_nphot_2014}. At high photon rates, specific output patterns occur (alternating, or coincidences), and the transition probability from one fixed output sequence is small. Therefore, the entropy per bit goes to zero (Fig.~\ref{fig:entropy}c). When coincidences are included, a third option ($\adet \bdet$) gives rise to a (binary) Shannon entropy above unity. Some timing information is introduced into this case, due to the fixed coincidence window. The calculated min-entropy stays in all cases below the curve for the Shannon entropy and below unity. 

Fig.~\ref{fig:entropy}d shows the total amount of produced entropy rate for the generator. As before, the calculated Shannon entropy exceeds all other curves, and including the coincidences allows for 25.1\% more extractable bits than a laser. All curves display an optimal point at which incident brightness the generator should be operated. 

The inset of Fig.~\ref{fig:entropy}d shows the optimum points of the min-entropy for the randomness generator. A single photon source peaks at a lower incident photon rate than the generator with a laser source. This is a global optimum and a laser, able to provide an ``arbitrary'' brightness, and no timing correlation, has lower usable entropy at all possible settings. One could imagine, that this point coincides with the count-rate advantage, presented in Fig.~\ref{fig:rawbit}b. The entropy peaks at an incident photon rate, which is 12.7\% higher than the peak in the rawbit rate -- not only the rawbit rate, but also the transition probabilities are relevant for the entropy. The pumping rate onto the single emitter, $k$, amounts to 1.1$\times$10$^8$~s$^{-1}$ (shown in Fig.~\ref{fig:setup}d).

The min-entropy is higher for a single photon source, but as for the laser source not all resulting rawbits can be used. When the min-entropy is known, this is the fraction of bits which can be extracted by a hashing procedure. This is ideally a two-universal hashing procedure~\cite{tomamichel_itoit_2011}. Other cryptographic hash functions, like SHA, are not designed or characterized for randomness extraction~\cite{young_book_2004,gabriel_np_2010}.

The described math underlines the advantages of bright single photon sources. The requirement is a collection efficiency close to unity, since the optimal point is around 50 million incident photons per second, whereas the maximum theoretical emission rate is $10^8$ photons per second. Such sources are experimentally realized~\cite{lee_np_2011,gazzano_nc_2013} and provide more than 45 $\times$ 10$^6$ photons per second. The presented results can be experimentally verified with such sources, assuming that the required speed of time-tagging electronics is available. Even if such sources present a background fraction, the extractable entropy remains comparable as to a ``perfect'' single photon source.

The described count-rate advantage is also relevant in microscopy applications of single emitters. Simply by the higher count-rates for a specific brightness, single photon emitters allow for a higher localization accuracy than a nanoscopic coherently emitting particle or when the radiative lifetime $T_1$ is very short. 
Very common, single emitters are researched with single photon detectors, but only by integrated detection. Since this characterizes the net brightness of the emitter, a relation to the real brightness is obscured when the count-rates are calculated equivalently to a laser source.

The given entropy enhancement is very robust in a wide range of input parameters. If the emitter lifetime is changed, the advantage stays in a range for $\mathit{\Gamma}$ of 1--18~ns. It is, though, optimal for an emitter lifetime of the given 10~ns. A crucial measure is the dead time of the detector which is assumed to be as measured in our case. A detector dead time of 30~ns shows still a small advantage for the detection of single photons. This is still better than the conservative bound of the present APD data-sheet (40~ns). The coincidence window also influences the outcome. With a shorter coincidence window, the advantage increases. The detector efficiency might be as low as 37\% and the beam splitter ratio can deviate from the ideal 50:50 case up to 70:30 -- still the described advantage holds.

It is evident that a true single photon source produces more usable entropy in an experimental configuration. A quantum random bit generator will benefit from bright single photon sources. Intrinsically, the non-classical nature of the source is measured, when the random bits are produced. This gives again some advantage against an external adversary, who might try to influence the generator's outcome~\cite{gerhardt_prl_2011,liu_rosi_2014}. The described phenomena also show, that an anti-bunched emitter can be more efficiently localized with the present single photon detectors than a Poisson emitter with the same brightness. The entropic advantage might enhance precision sensing and microscopic applications in single emitter studies even further. How far this is possible is presently ongoing research.

\bibliographystyle{unsrt}
\bibliography{../../references/trng01}

\section*{Acknowledgments}
Axel Kuhn provided some inputs on the statistical description of single photon measurements. We acknowledge the critical questions and ideas provided by Daniela Frauchiger and Renato Renner. We thank J\"org Wrachtrup for continuous support.

\pagebreak
\widetext
\clearpage
\begin{center}
\textbf{\large Supplementary material to:\\Better Randomness with Single Photons}
\end{center}
\setcounter{equation}{0}
\setcounter{figure}{0}
\setcounter{table}{0}
\setcounter{page}{1}
\makeatletter
\renewcommand{\theequation}{S\arabic{equation}}
\renewcommand{\thefigure}{S\arabic{figure}}
\renewcommand{\bibnumfmt}[1]{[S#1]}
\renewcommand{\citenumfont}[1]{S#1}

\newcommand{\ket}[1]{|{#1}\rangle}
\newcommand{\bra}[1]{\langle{#1}|}
\newcommand{\braket}[1]{\langle{#1}\rangle}

\newcommand{\renyiorder}{\mathbold{\alpha}}

\newcommand{\sps}{single photon source}

\addtocounter{figure}{4}

\tableofcontents

\section{Overview}

In this supplement we introduce the steps to derive the entropy of the introduced QRNG.

We follow the following procedure to derive the final entropy of the generator.

\begin{enumerate}
\item describe the generator and the possible detector outcomes. This includes assumptions on the detection process, external adversaries, etc.. All detector outcomes can be described in this model.
\item define the raw generator outcomes. Here we define which outcomes of the generator are resulting in what outcomes. We call this entity the ``raw-bit'' rate, although we partially derive three outcomes.
\item describe all probabilities and conditional transition probabilities in the system. This is required to judge if there are correlations in the system.
\item calculate the conditional entropies in the system. Afterwards we know how much entropy can be extracted.
\item post process the raw-bits according to the entropy-fraction of the generator
\end{enumerate}

This scheme can be generalized to many random number generators. We recommend the study of~\cite{frauchiger_a_2013} to extend our approach presented below.

\section{The single photon source and the $g^{(2)}$-function}

\begin{figure}[H]
\begin{center}
\includegraphics[width= 0.9 \textwidth]{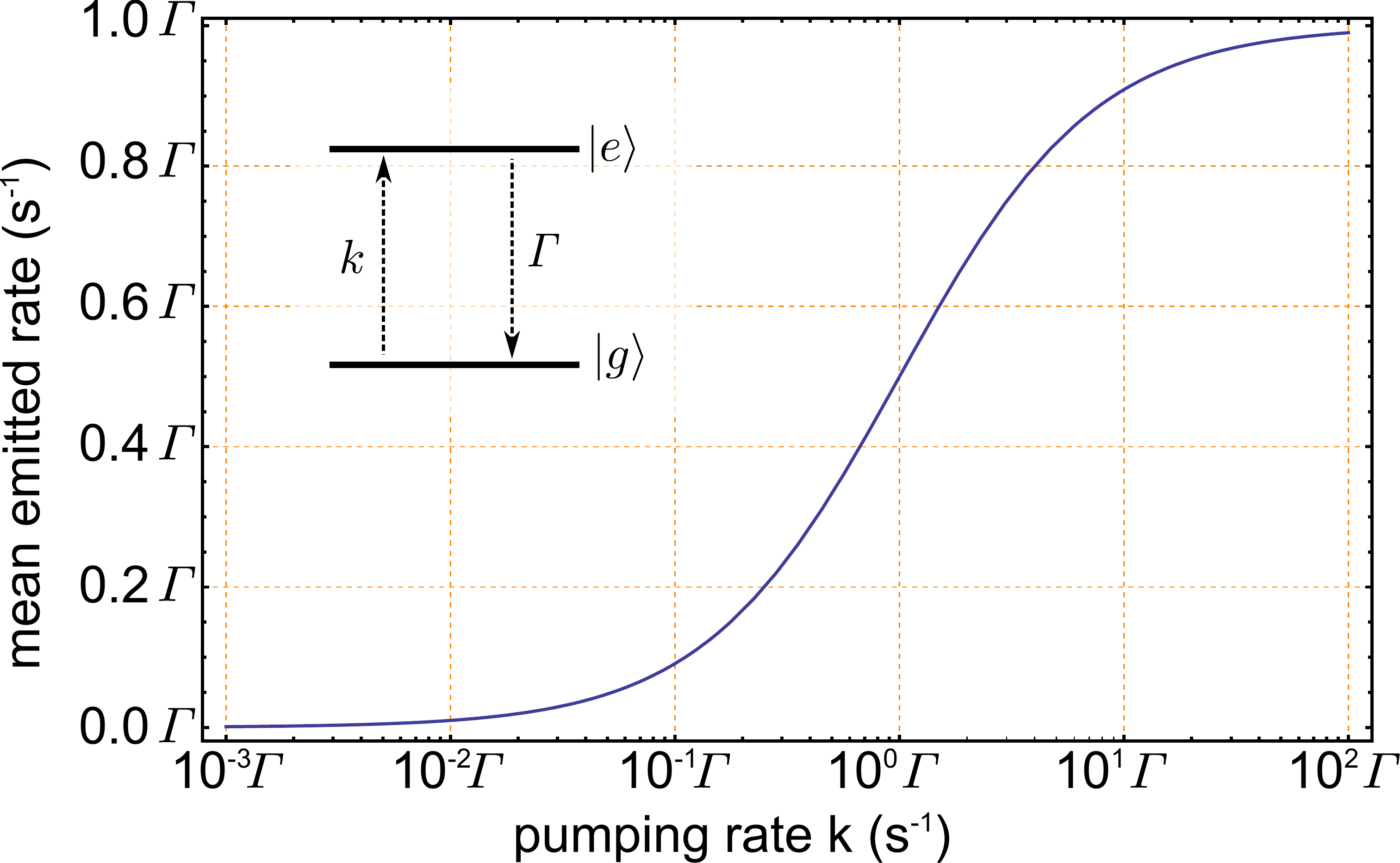}
\caption{The saturation curve $\lambda(k)$ gives the rate of emitted photons by the single photon source. It depends on the intensity of the pumping laser. The saturation value for high pumping rates $k$ is $\lambda = \mathit{\Gamma}$. Inset: shows the two level system. Since the transition from a higher laying excited state to $\ket e$ is assumed to happen instantaneously, the transition from $\ket{g}$ over higher level to $\ket e$ is modeled direct from $\ket g$ to $\ket e$.}
\label{fig:sps_saturation}
\end{center}
\end{figure}

The single photon source, which is regarded in this paper, is based on a single emitter. The working scheme is given in the inset of Fig.~\ref{fig:sps_saturation}. It is equivalent to many reports in the literature~\cite{kimble_prl_1977,diedrich_prl_1987,basche_prl_1992,lounis_n_2000}, when the coherences are neglected~\cite{lounis_n_2000,orrit_sm_2002}.

A pump rate $k$ brings the emitter from ground state $\ket g$ to an excited state $\ket {e}$. The single photons originate from the transition $\ket e$ to $\ket g$. The decay constant for the latter is $\mathit{\Gamma}$. The mathematical description of the change in the population $\rho_{11}$ ($\rho_{22}$) of state $\ket g$ ($\ket e$) is given by Eqn.~\ref{eq:dglsyst_sps_g} (\ref{eq:dglsyst_sps_e}).

\begin{eqnarray}
\label{eq:dglsyst_sps_g}
\dot{\rho}_{11}&=&-k\rho_{11} + \mathit{\Gamma} \rho_{22}\\
\label{eq:dglsyst_sps_e}
\dot{\rho}_{22}&=&+k\rho_{11} - \mathit{\Gamma} \rho_{22} \ ,
\end{eqnarray}

\noindent
This differential equations are solved for a single emitter, assuming the emission of a single photon at $t=0$ ($\rho_{11}(0)=1$ and $\rho_{22}(0)=0$). The solution denotes

\begin{eqnarray}
\label{eq:rho_g_sps}
\rho_{11}(\tau)&=& \frac{k\cdot e^{-(k+\mathit{\Gamma})\tau}+\mathit{\Gamma}}{k+\mathit{\Gamma}} \\
\label{eq:rho_e_sps}
\rho_{22}(\tau)&=& \frac{k \left(1 - e^{-(k+\mathit{\Gamma})\tau} \right) }{k+\mathit{\Gamma}} \ .
\end{eqnarray}

\noindent
It is evident in Eqn.~\ref{eq:rho_g_sps} and \ref{eq:rho_e_sps}, that the system goes for a constant value of $k$ into an equilibrium, as $\tau \rightarrow \infty$. In this state of equilibrium the emitted photon rate $\lambda$ of the single photon source denotes

\begin{equation}
\label{eq:lambda_sps}
\lambda(k)=\mathit{\Gamma}\cdot \lim_{\tau\rightarrow\infty}{\rho_{22}(\tau)}=\frac{k\mathit{\Gamma}}{k+\mathit{\Gamma}} \ .
\end{equation}
$\lambda$ as a function of $k$ is shown in Fig.~\ref{fig:sps_saturation}.

For $k\rightarrow \infty$ the emitted photon rate goes to $\mathit{\Gamma}$. The population of the excited state is $\rho_{22}\rightarrow 1>0.5$ (population inversion). The property, that the single photon source never emits more than one photon at the same time, can be observed by calculating the $g^{(2)}$ correlation function of the source.

It is defined as~\cite{fox_book_2006}

\begin{equation}
g^{(2)}(\tau)=\frac{\braket{I(t)I(t+\tau)}}{\braket{I(t)} \braket{I(t+\tau)}}
\label{eq:g2}
\end{equation}

\noindent
and in the denormalized form

\begin{equation}
G^{(2)}(\tau)=\braket{I(t)I(t+\tau)} \ ,
\label{eq:G2}
\end{equation}

\noindent
where $I$ describes the intensity of the light source. If we treat light as quantized particles -- photons -- $I$ describes the emitted photon rate and the $g^{(2)}$-function gives the statistics that the source emit a photon at time $t+\tau$ under the condition that the light source has emitted a photon at time $t$. If the light source, attenuated by the parameter $\eta$ (with $\eta \in (0,1]$), goes onto a perfect detector, the $g^{(2)}$-function describes also the detection statistics of the detector. In the following the parameter $\eta$ is included into the intensity $I$ of the light source, that means $I$ is the attenuated intensity of the light source.\\
A simple setup to measure approximately the $g^{(2)}$-function can be described by the same setup as our random number generator uses (see Fig.~\ref{fig:setup}a). It has to be performed as a start-stop measurement. This measurement only results in the statistics about the waiting time to the detection of the \emph{next} photon, but if the light source is attenuated strongly, this statistics is for short enough times a good approximation to the $g^{(2)}$-function. 

A helpful property of the correlation-function is the invariance under time inversion, that means $g^{2}(\tau)=g^{2}(-\tau)$. There is no difference between measuring the time from the start event to stop event or vice versa. The physics is the same.

A stable laser with a fixed intensity $I_0$, attenuated by $\eta$, has for each point in time the correlation-function

\begin{equation}
g^{(2)}_{\mathrm{laser}}(\tau)=\frac{(\eta I_0)^2}{(\eta I_0)^2}=1 \quad \forall \, \tau \ .
\label{eq:g2_laser}
\end{equation} 

\noindent
A chunk of radioactive material follows the same correlation-function. There are no correlations among different decay events. One decay does not influence the next (or coinciding) outcome. Equivalently, the emission of a photon has no effect on other photons. The denormalized correlation function of a laser which is attenuated by the factor $\eta$ denotes

\begin{equation}
G^{(2)}_{\mathrm{laser}}(\tau)= (\eta \lambda)^2 \ ,
\label{eq:G2_laser}
\end{equation} 

\noindent
where $\lambda$ is the mean emitted photon rate of the laser.\\

To derive the correlation function of the single photon source the population of the excited state $\ket e$ needs to be known. These population $\rho_{22}(\tau)$ can be understood as the probability that the system is in the excited state $\ket e$ at time $\tau$, which is proportional to the photon emission probability at time $\tau$. The normalized correlation function of the single photon source reads

\begin{equation}
g^{(2)}_{\mathrm{sps}}(\tau)	= \frac{\rho_{22}(|\tau|)}{\lim_{\tau \rightarrow \infty} {\rho_{22}(\tau)}}
= 1-e^{-(k+\mathit{\Gamma})|\tau|} \ .
\label{eq:g2_sps}
\end{equation}

\noindent
The denormalized correlation function of the single photon source, attenuated by the factor $\eta$, is

\begin{equation}
G_{\mathrm{sps}}^{(2)}(\tau)= (\eta \underbrace{ \lambda(k) }_{\frac{k \mathit{\Gamma}}{k+\mathit{\Gamma}}} )^2 \left( 1-e^{-(k+\mathit{\Gamma})|\tau|} \right) \ .
\label{eq:G2_sps}
\end{equation}

\noindent
Fig.~\ref{fig:setup}d shows the correlation functions $g^{(2)}(\tau)$ of the laser and the single photon source for two values of the pumping rate $k$. The so-called ``anti-bunching dip'' gets more and more narrow with higher pumping rates.

The property of the single photon source, never to emit more than one photon at the same time, is fulfilled because of $g_{\mathrm{sps}}^{(2)}(\tau=0)=0$. Due to this property this light is called ``anti-bunched''.

\section{$JK$-formalism}


The $g^{(2)}(\tau)$-function tells only about the detection statistics of \emph{any} photon at time $\tau$, under the condition, that at time $t=0$ a photon has been detected. With the $JK$-formalism~\cite{reynaud_a_1983}, the probability density of detecting exactly the \emph{next} photon after the 0th photon at $t=0$ can be predicted. The function $K(\tau)$ is the probability density of the \emph{next} photon and the function $J(\tau) := g^{(2)}(\tau)\cdot(\eta \lambda)$ describes the number density for the detection of \emph{any} photon at time $\tau$. The functions $L_m(\tau)$ are defined as the probability density for the arrival of the $m$th photon (with $m \in \mathbb{N}$), conditioned on the $0$th photon at $t=0$. It is

\begin{equation}
\sum_{m=1}^\infty L_m(\tau)=J(\tau) \ .
\label{eq:sum_Lm}
\end{equation}

\noindent
The probability density of the $(m+1)$th photon can be predicted through the probability density of the $m$th photon and the function $K(\tau)$:

\begin{equation}
L_{m+1}(\tau)=\int_{0}^{\tau}L_{m}(\tau-t)K(t)\ dt \ = (L_m \ast K) (\tau) \quad \mathrm{with} \quad L_1(\tau)=K(\tau) \ ,
\label{eq:L_m+1}
\end{equation}

\noindent
where $\ast$ denotes the convolution. If the function $K(\tau)$ is known, all functions $L_m$ can be predicted in an iterative through Eqn.~\ref{eq:L_m+1}. $K$ can be derived by

\begin{subequations}
\label{eq:derive_K}
\begin{align}
							\int_{0}^{\tau}J(\tau-t)K(t)\ dt &= \sum_{m=2}^\infty L_m(\tau) \\ 
\Leftrightarrow \qquad		(J*K)(\tau) &= J(\tau)-K(\tau) \\
\label{eq:derive_K_c}
\Leftrightarrow \qquad		\mathcal{L}[(J*K)(\tau)] &= \mathcal{L}[J(\tau)]-\mathcal{L}[K(\tau)] \\
\Leftrightarrow \qquad		\mathcal{L}[J(\tau)] \cdot \mathcal{L}[K(\tau)] &= \mathcal{L}[J(\tau)]-\mathcal{L}[K(\tau)]\\
\label{eq:K}
\Leftrightarrow \qquad K(\tau) &= \mathcal{L}^{-1} \left[ \frac{\mathcal{L}[J(\tau)]}{1+\mathcal{L}[J(\tau)]} \right] \ ,
\end{align} 
\end{subequations}

\noindent
where where $\mathcal{L}[f]$ denotes the Laplace-transformation of the function $f$. The term on the left in Eqn.~\ref{eq:derive_K_c} is simplified by the convolution theorem~\cite{bronstein_book}.

\section{Arrival time of the $m$th photon}
To study the emission behavior of the light sources in more detail than the $g^{(2)}$-function tells us, the probability densities $L_m(\tau)$ of the $m$th photon after the $0$th photon at $t=0$ will be estimated. With Eqn.~\ref{eq:K} and the number densities
\begin{eqnarray}
J_{\mathrm{laser}}(\tau)&=&	\eta \lambda \\
J_{\mathrm{sps}}(\tau)&=&		\eta \lambda \left( 1- e^{-(k+\mathit{\Gamma})t} \right)
\end{eqnarray}

\noindent
the probability density $K(\tau)=:L_1(\tau)$ of the next photon, i.e. the first ($m$=1), is predicted for the laser and the \sps. The resulting functions are

\begin{eqnarray}
&&K_{\mathrm{laser}}(\tau)= \eta \lambda \cdot e^{-\eta \lambda \tau} \\
&&K_{\mathrm{sps}}(\tau)= \frac{2 \eta \lambda \sqrt{k+\mathit{\Gamma}}}{\sqrt{k+\mathit{\Gamma}-4\eta\lambda}} \cdot
e^{-\frac{1}{2}(k+\mathit{\Gamma})\tau} \cdot
\sinh{\left( \frac{1}{2} \sqrt{k+\mathit{\Gamma}} \sqrt{k+\mathit{\Gamma}-4\eta\lambda} \right)} \ .
\end{eqnarray}

\noindent
With Eqn.~\ref{eq:L_m+1} the probability densities $L_{m}(\tau)$ of the $m$th photon can be calculated iteratively. For the laser an analytic expression is given by

\begin{equation}
L_{m, \mathrm laser}(\tau)=\frac{(\eta\lambda)^m \tau^{m-1}}{(m-1)!} e^{-\eta\lambda\tau} \quad \mathrm{with}\ m \in \mathbb{N} \ .
\label{eq:L_m_laser}
\end{equation}

\noindent
The functions in the case of the \sps ~become very lengthy. The \textsc{Mathematica} source code of the calculation can be obtained in Appendix~\ref{sect:Lm_Pm}.


\noindent
The probability densities $L_m(\tau)$ are shown in Fig.~\ref{fig:setup}e for both light sources with the parameters $\lambda=10^7~{\mathrm s}^{-1}$, $\mathit{\Gamma}=10^8~\mathrm s^{-1}$ and $\eta=1$.

The greatest difference between the two sources exists between the densities of the first photon. Exactly this functions give a strong hint, that the \sps ~produces less coincidences than the laser, because for small waiting times ($\tau_{\mathrm{cw}}=2$~ns) it is more unlikely that the \sps ~emits the first photon.

\section{Number of photons within time intervals}

The described single photon source is a ``real'' single photon source, when it is operated with pulsed excitation, by e.g. a pulsed laser. If the laser has sufficient energy, the system is transferred to the excited state, and emits only a single photon. When the device is operated in continuous wave (cw) mode, the source emits only a single photon at a time -- but for a finite observation window, more photons could arrive on a detector. Therefore, we first calculate the probability distribution for a certain number of photons within a waiting time $\tau$. The function $P_m(\tau)$ (with $m \in \mathbb{N}_0$) gives the probability that $m$ photons will occur within the time interval $[0,\tau]$ under the condition, that the 0th photon was at $t=0$. Note, that the 0th photon is not included into the number $m$. These functions can be predicted through 

\begin{eqnarray}
P_0(\tau)&=&1-\int_{t=0}^\tau K(t)\, \mathrm{d}t \\
P_m(\tau)&=	&\int_{t'=0}^\tau \left[ L_m(t')
\underbrace{ \left( 1- \int_{t=0}^{\tau-t'}K(t)\ dt \right) }_{P_0(\tau-t')} \right] \, \mathrm{d}t' 
\quad \mathrm{with} \ m \in \mathbb{N} \ .
\end{eqnarray}

\noindent
The integral $\int_{t=0}^\tau K(t)\, \mathrm{d}t$ is the probability that the \emph{next} photon occur in the interval $[0,\tau]$. The probability $P_0(\tau)$ is calculated by the probability, that the \emph{next} photon will not come within $[0,\tau]$. To calculate $P_m(\tau)$ the probability has to be determined, that at $t=t' \in [0,\tau]$, the $m$th Photon has occurred ($L_m(t')$) \emph{and} within the remaining time $\tau-t'$, there will not come the following photon ($P_0(\tau-t')$). At a later date we need the probability $P_0([t_1,t_2])$ that within a time interval $[t_1,t_2]$ (with $(t_1,t_2 > 0) \ \land \ (t_1<t_2)$ ) no photon will be emitted, when at $t=0$ a photon was emitted. This probability is defined as:

\begin{equation}
\label{eq:P_0(t1,t2)}
P_0(t_2)=P_0(t_1) \cdot P_0([t_1,t_2]) \quad \Leftrightarrow \quad P_0([t_1,t_2])=\frac{P_0(t_2)}{P_0(t_1)}
\end{equation}

\noindent
In the case of the laser the functions $P_m(\tau)$ denote

\begin{equation}
\label{eq:P_m,laser}
P_{m, \mathrm laser}(\tau) 	= \frac{(\eta\lambda \tau)^m}{m!} \cdot e^{-\eta\lambda \tau}
=: \frac{\braket{m}^m}{m!} e^{-\braket{m}} = \text{Poi} (m)
\quad \mathrm with \ m \in \mathbb{N}_0 \ ,
\end{equation}

\noindent
where $\braket m := \eta\lambda \tau$ denotes the expected value of number $m$. Eqn.~\ref{eq:P_m,laser} can be identified with the Poisson distribution $\mathrm{Poi}(m)$. Of course, this was known for a laser. But a very interesting point is, that the 0th photon at $t=0$ is not involved in this Poisson distribution. If at $t=0$ there would be no photon we would expect exactly the same distribution. The fact, that an detected photon, which is then not included in the distribution, does not change the photon emission distribution of the laser, was described before~\cite{kuhn_t_2005}. The proof can be outlined as follows:\\
The detection of a photon means a transition from the occupancy state $\ket m$ to $\ket{m-1}$. The probability $P(m-1)$ to be after a detection in the state $\ket{m-1}$ is proportional to the probability $P(m)$, that the detected photon originate from the state $\ket m$:

\begin{equation}
\label{eq:P(m-1)}
P(m-1) \propto P(m)=\binom{m}{1}\text{Poi}(m)=m\cdot \text{Poi}(m)\ .
\end{equation}

\noindent
Eqn.~\ref{eq:P(m-1)} can be prepared by the substitution $k\mapsto k+1$:

\begin{subequations}
\begin{eqnarray}
P(m) &\propto & (m+1) \cdot \text{Poi}(m+1) \\
 &=& (m+1) \frac{\braket{m}^{m+1}}{(m+1)!}e^{-\braket{m}} \\
 &=& \braket{m} \cdot \frac{\braket{m}^m}{m!}e^{-\braket{m}} \\
 &=& \braket{m} \cdot \text{Poi}(m)
\end{eqnarray}
\end{subequations}

\noindent
Now we got the expression $P(m)\propto \braket{m} \cdot \text{Poi}(m)$, which has to be normalized by the norm $N$:

\begin{equation*}
\frac{1}{N} \braket{m} \underbrace{\sum_{m=0}^\infty \text{Poi}(m)}_{=1} \overset{!}{=} 1 \quad \Rightarrow \quad N=\braket{m}
\end{equation*} 

\noindent
The new distribution $P(m)$ after the detection of a photon is still the Poisson distribution: $P(m)=\text{Poi}(m)$. \\
The calculated functions $P_m(\tau)$ in the case of the \sps ~become also very lengthy. The \textsc{Mathematica} source code of the calculation can be obtained in Appendix~\ref{sect:Lm_Pm}.\\
For the same parameters as above ($\lambda=10^7~{\mathrm s}^{-1}$, $\mathit{\Gamma}=10^8~\mathrm s^{-1}$ and $\eta=1$), the probabilities $P_m(\tau)$ are shown for some different values of $\tau$ in Fig.~\ref{fig:setup}f for both light sources.

It is more probable that no further photons come within the waiting time $\tau$ for the \sps: $P_0(\tau)_{\mathrm{sps}}>P_0(\tau)_{\mathrm laser}$. In the case of the laser the probability distribution $P_{m,\mathrm laser}(\tau)$ is given through the Poisson distribution $\mathrm{Poi}_m(\braket{m})=\braket{m}^m/(m!)\cdot \exp{(-\braket{m})}$ with the parameter $\braket{m}=\eta\lambda\tau=1\cdot 10^7~\mathrm{s}^{-1}\cdot \tau$. 

\section{Click rate of the detectors}

Now the focus goes to the random number generator. In this section the click rate of each detector, which is limited through the dead time $\tau_{\mathrm{dead}}$, will be estimated.\\
But first it has to be explained, how to adapt all previous defined math to the specific problem. What has to be included is the beam splitter ratio and the quantum efficiency of the APDs. All these defined functions above can be transferred to the case of the detector $\adet \in \{R,T\}$ by simply substitute the attenuation $\eta \mapsto \eta^\adet:=\eta_{\mathrm{\mathrm{qe}}} p_\adet$, where $\eta_{\mathrm{\mathrm{qe}}}$ denotes the quantum efficiency and $p_\adet$ the probability that the photon will take on the beam splitter the path to detector $\adet$. In principle, these defined attenuation can be interpreted as a grey-filter, which attenuates the emitted light, which then impinges onto a perfect detector $\adet$ with a quantum efficiency of unity (but still the influence of dead time has to be considered). The new function will be labeled by an upper index $\adet$ (for example $J^\adet(\tau)$ is the number density of the photons which ``decided'' on the beam splitter the $\adet$-path and are definitely not overseen by detector $\adet$ because of the quantum efficiency).\\

The click rate $\lambda_{\mathrm{click}}^{\adet}$ of detector $\adet$ has to be calculated. The limiting factor is the dead time $\tau_{\mathrm{dead}}=50~\mathrm ns$ of the detectors. First, we determine the average number $m^\adet$ of photons which occur within a dead time on detector $\adet$

\begin{equation}
m^\adet=\int_0^{\tau_{\mathrm{dead}}} J^{\adet}(\tau) \ d\tau \ .
\label{eq:m_alpha}
\end{equation}

\noindent
One click on detector $\adet$ leads to one dead time of length $\tau_{\mathrm{dead}}$. That means that from these $\lambda^\adet:=\lambda \eta^\adet$ incoming photons to detector $\adet$, which definitely produce a click, if the detector is not in dead time, $\lambda_{\mathrm{click}}^{\adet}$ photons will produce a ``click'' and $\lambda_{\mathrm{click}}^{\adet} \cdot m^\adet$ of these photons will enter while the detector is blind. So the click rate $\lambda_{\mathrm{click}}^{\adet}$ is

\begin{equation}
\lambda^\adet= \lambda_{\mathrm{click}}^{\adet} (1+m^\adet) \qquad \Leftrightarrow \qquad \lambda_{\mathrm{click}}^{\adet}=\frac{\lambda^{\adet}}{1+m^\adet} \ .
\label{eq:lambda_2}
\end{equation}

\begin{figure}[H]
\begin{center}
\includegraphics[width=0.9 \columnwidth]{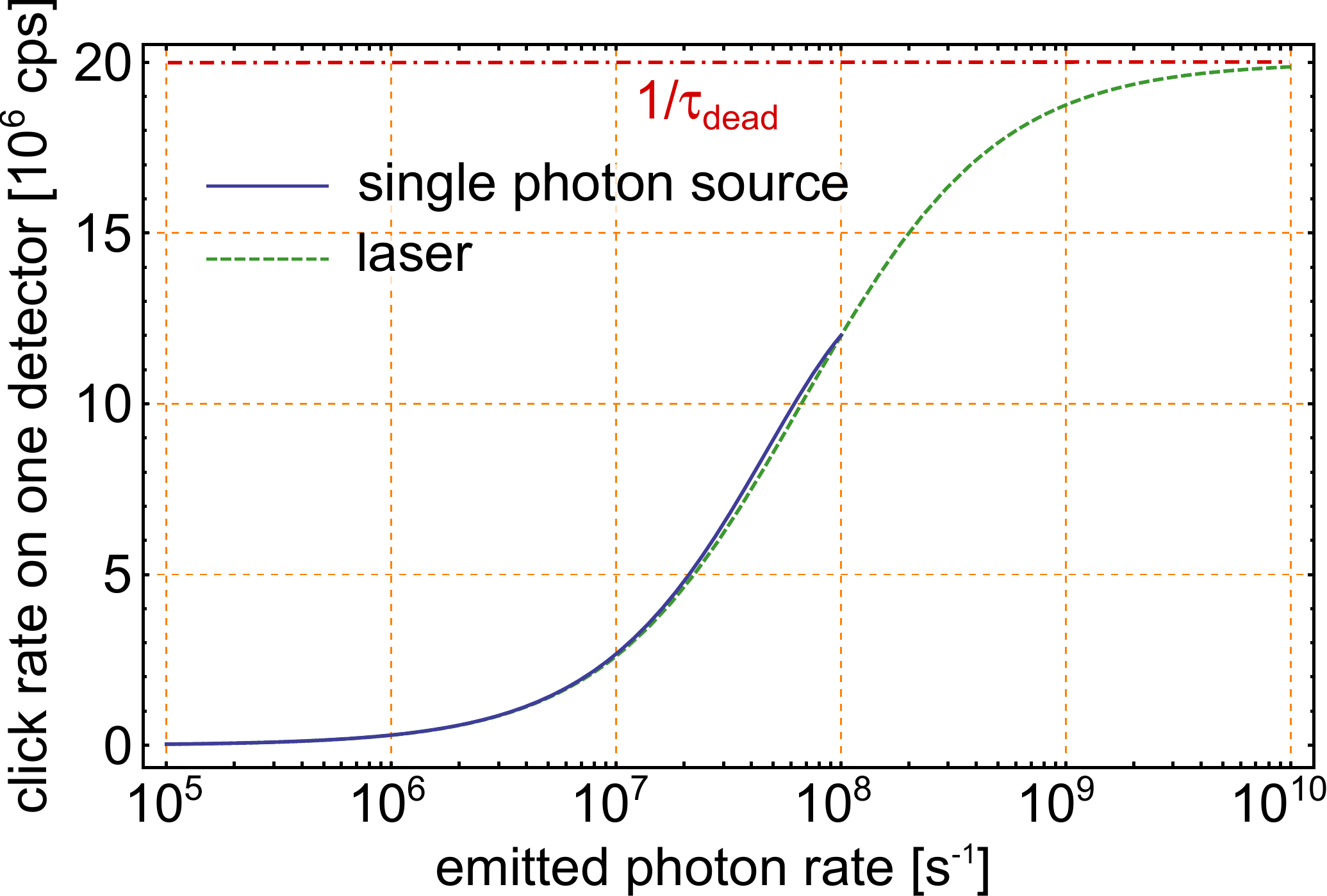}
\caption{The click rate $\lambda^R_{\mathrm{click}}$ of detector \texttt{R} is plotted for the parameters
$\eta_{\mathrm{qe}}=0.6$, $p_R=0.5$, $\mathit{\Gamma}=10^8~\mathrm s^{-1}$ and $\tau_{\mathrm{dead}}=50$ ns. The single photon source has a higher click rate. The click rate of the laser converges to the click rate $1/\tau_{\mathrm{dead}}=2 \times 10^7$~s$^{-1}$ for high laser intensities.}
\label{Fig:clickrates}
\end{center}
\end{figure}

\noindent
The click rates of detector \texttt{R} or \texttt{T} (since $p_R=p_T=0.5$) are shown in Fig.~\ref{Fig:clickrates} for the laser and the single photon source.

It can be seen, that the single photon source produces more detector clicks than the laser until to the maximal emission rate ($\lambda_\mathrm{max,\, sps}=\mathit{\Gamma}=10^8$~s$^{-1}$) of the single photon source. But the laser can be driven at higher photon rates and as the intensity is increased, the click rate converges to the maximal achievable click rate per detector ($1/\tau_{\mathrm{dead}}$).

\section{Raw bit rate}

Now the question arises, whether a click on detector $\adet$ at $t=0$ produces a bit or not, because beneath the dead time another technical influence is the coincidence window (with length $\tau_{\mathrm{cw}} = 2$~ns). When detector $\adet$ clicks, there are two different scenarios which could happen on detector $\bdet$ (with $\bdet \in \{R,T\} \land \bdet \neq \adet$):

\begin{itemize}
\item detector $\bdet$ clicks within the time interval $[-\tau_{\mathrm{cw}},+\tau_{\mathrm{cw}}]$\\
$\Rightarrow$ coincidence\\
$\Rightarrow$ we discard this output.
\item detector $\bdet$ clicks not within the time interval $[-\tau_{\mathrm{cw}},+\tau_{\mathrm{cw}}]$\\
$\Rightarrow$ a bit is produced by detector $\adet$.
\end{itemize}

In the following we calculate the bit rate of detector $\adet$ ($\lambda_{\mathrm{click}}^\adet$) and the coincidence rate ($\lambda_{\mathrm{coinc}})$. To that end it has to be calculated the probability, that detector $\bdet$ clicks within the time interval $[-\tau_{\mathrm{cw}},\tau_{\mathrm{cw}}]$, conditioned on a click of detector $\adet$ at time $t=0$. It is not sufficient to regard whether within the time interval $[-\tau_{\mathrm{cw}},\tau_{\mathrm{cw}}]$ a photon will enter detector $\bdet$, because the latter might be in dead time. Instead we make use of the above calculated click rate of detector $\bdet$. We know that the average number of clicks within a time interval of length $2\tau_{\mathrm{cw}}<\tau_{\mathrm{dead}}$ reads

\begin{equation}
\braket{\lambda_{\mathrm{click}}^\bdet} \big|_{[-\tau_{\mathrm{cw}},\tau_{\mathrm{cw}}]} = 2 \tau_{\mathrm{cw}} \lambda_{\mathrm{click}}^\bdet \ .
\end{equation}

\noindent
This quantity can also be interpreted as the probability for a click in the interval, because the interval length is shorter than one dead time an therefore the only possibilities are zero or one click. The problem changes (in the case of the single photon source) when we postulate a click at $t=0$ on detector $\adet$ and regard the time interval $[-\tau_{\mathrm{cw}},\tau_{\mathrm{cw}}]$ on detector $\bdet$. Due to the collapse of the $g^{(2)}$-function at $t=0$ (in the case of the single photon source) the probability that detector $\bdet$ clicks decrease. The click probability density of detector $\bdet$ reads then 

\begin{equation}
\rho_{\mathrm{click}}(\tau) = \braket{\lambda_{\mathrm{click}}^\bdet} \big|_{[-\tau_{\mathrm{cw}},\tau_{\mathrm{cw}}]} \cdot \frac{ g^{(2)} (\tau) }{2 \tau_{\mathrm{cw}}} =\lambda_{\mathrm{click}}^\bdet g^{(2)}(\tau) \ .
\end{equation}

\noindent
The probability $P_{\mathrm{coinc}}^\bdet$, that detector $\bdet$ clicks within the interval $[-\tau_{\mathrm{cw}},\tau_{\mathrm{cw}}]$ when there is a click at $t=0$ on detector $\adet$, which means that a coincidence occurs, is

\begin{equation}
P_{\mathrm{coinc}}^\bdet	= \int_{-\tau_{\mathrm{cw}}}^{\tau_{\mathrm{cw}}} \rho_{\mathrm{click}}(\tau) \, \mathrm{d}\tau
= \lambda_{\mathrm{click}}^\bdet \int_{-\tau_{\mathrm{cw}}}^{\tau_{\mathrm{cw}}} g^{(2)}(\tau) \, \mathrm{d}\tau
=2 \cdot \lambda_{\mathrm{click}}^\bdet \int_{0}^{\tau_{\mathrm{cw}}} g^{(2)}(\tau) \, \mathrm{d}\tau \ .
\label{eq:P_coinc}
\end{equation}

\noindent
Note, that Eqn.~\ref{eq:P_coinc} diverges for $\tau_{\mathrm{cw}} \rightarrow \infty$. But a probability should always be smaller than one. This equation is only an approximation for small coincidence windows (which is given in our considerations)!\\
Now we are able to estimate the coincidence rate ($\lambda_{\mathrm{coinc}}=:\lambda_{\mathrm{bit}}^{\adet \bdet}$)

\begin{equation}
\lambda_{\mathrm{bit}}^{\adet \bdet}:=\lambda_{\mathrm{coinc}} =	\lambda_{\mathrm{click}}^\adet P^\bdet_{\mathrm{coinc}}=
\lambda_{\mathrm{click}}^\bdet P^\adet_{\mathrm{coinc}}
\end{equation}

\noindent
and the bit rate of the detector $\adet$ ($\lambda_{\mathrm{bit}}^\adet$)

\begin{equation}
\lambda_{\mathrm{bit}}^\adet = \lambda_{\mathrm{click}}^\adet \left( 1- P^\bdet_{\mathrm{coinc}} \right) \ .
\end{equation}
												
\noindent
The whole bit rate $\lambda_{\mathrm{bit}}$ is the sum over both detectors:

\begin{equation}
\lambda_{\mathrm{bit}} = \lambda_{\mathrm{bit}}^\adet + \lambda_{\mathrm{bit}}^\bdet
= \lambda_{\mathrm{click}}^\adet + \lambda_{\mathrm{click}}^\bdet -2 \lambda_{\mathrm{coinc}} \ .
\end{equation}

Fig.~\ref{fig:rawbit}a depicts the bit rate $\lambda^R_{\mathrm{bit}}$ of detector \texttt{R} (or \texttt{T}, since $p_R=p_T=0.5$) for some different parameters $\tau_{\mathrm{cw}}$ and also the coincidence rate $\lambda_{\mathrm{coinc}}$ for the standard parameter $\tau_{\mathrm{cw}}=2$~ns.

Note, that the bit rate for the parameter $\tau_{\mathrm{cw}}=0$ is exactly the click rate as shown in Fig.~\ref{Fig:clickrates}, since there exist no coincidences and every click automatically becomes a bit. As the parameter $\tau_{\mathrm{cw}}$ increases, the bit rate decreases (and thus the coincidence rate increase). But for each value of $\tau_{\mathrm{cw}}$ the single photon source produces more raw bits within its emission range. On the bottom can be seen, that the single photon source has less coincidences than the laser. The purple function shows the bit rate for $\tau_{\mathrm{cw}}=0$~ns, i.e. the click rate, if the beam splitter is taken out of the setup and all light impinges onto only one detector. This has the effect, that for small incoming photon rates the click rate becomes double compared to the case with beam splitter, and for high photon rates the click rate converges faster against $1/\tau_{\mathrm{dead}}$.\\

In Fig.~\ref{fig:rawbit}b is shown the function $(\lambda_{\mathrm{bit,\, sps}}^\adet-\lambda_{\mathrm{bit,\, laser}}^\adet)/\lambda_{\mathrm{bit,\, laser}}^\adet$, which gives the of enhancement of the raw bit rate when using the single photon source instead of the laser, for some parameters $\tau_{\mathrm{cw}}$.

When discarding coincidences, the more the coincidence window $\tau_{\mathrm{cw}}$ is increased, the more is the single photon source superior against the laser. But even if the coincidence window is zero, the single photon source produces on the maximum point 4.7\% more bits. With the standard parameter $\tau_{\mathrm{cw}}=2$~ns the enhancement denotes on its maximum 7.5\%. The enhancement of the single photon source in the setup without beam splitter, where all light impinges onto one detector is shown by the purple curve. 

\section{Entropy}

The entropy $\mathcal{H}$ of a random number is the amount of information, the number is containing. It is important to know these quantity of a raw bit string, to extract the maximal information of the string in a subsequent post-processing step.

The R\'{e}nyi entropy~\cite{renyi_proc_1961} of order $\renyiorder$ (with $\renyiorder > 0$ and $\renyiorder \ne 1$) is defined as 

\begin{equation}
\mathcal{H}_\renyiorder(X)= \frac{1}{1-\renyiorder} \log_2{ \left[ \sum_{x=1}^{n} p(x)^\renyiorder \right]} \ ,
\label{eq:renyi}
\end{equation} 

where $X$ is a random variable with the outcomes $x$ (with $x \in \{1,2,...,n\}$). Please note that this definition of $\renyiorder$ contradicts with the definition for the coherent state ($\alpha$) in the main part of this paper. Therefore, we utilize a bold version at this point. The $x$th outcome occurs with the probability $p(x)$, while $\sum_{x=1}^n p(x)=1$. The limit case $\renyiorder \rightarrow 1$ gives the Shannon entropy
\begin{equation}
\mathcal{H}_{\mathrm Shannon}(X)	=\lim_{\renyiorder \rightarrow 1} \mathcal{H}_\renyiorder(X)
							= -\sum_{x=1}^n p(x) \log_2 \left[ p(x) \right] \ .
\label{eq:shannon}
\end{equation}
And the limit case $\renyiorder \rightarrow \infty$ gives the min-entropy
\begin{eqnarray}
\label{eq:min}
\mathcal{H}_{\mathrm min}(X)	&=&\lim_{\renyiorder \rightarrow \infty} \mathcal{H}_\renyiorder(X)
=-\log_2{ \left[ \mathrm{Guess}(X) \right] } \quad ,~\mathrm{with}\\ \notag
\mathrm{Guess}(X) &=& \max_{x}\{p(x)\}
\end{eqnarray}

\noindent
The min-entropy is the smallest R\'{e}nyi entropy. For our application it is the most important quantity, since it is an conservative bound on the entropy fraction in the raw bit stream.\\
But it might be, that the outcome of the random variable $Y$ has an influence on the outcome of the random variable $X$. To describe the entropy of such a coupled system, a conditional entropy is required~\cite{fehr_online_2013}. The conditional Shannon entropy of the random variable $X$ conditioned on the random variable $Y$ is
\begin{eqnarray}
\label{eq:cond_shannon}
\mathcal{H}_{\mathrm Shannon}(X|Y)	&=& \sum_y p(y) \mathcal{H}_{\mathrm Shannon}(X|Y=y)\\ \notag
&=&-\sum_y p(y) \sum_x p(x|y) \log_2{\left[ p(x|y) \right]}
\label{eqn:shan}
\end{eqnarray} 

\noindent

and conditional min-entropy

\begin{eqnarray}
\mathcal{H}_{\mathrm min}(X|Y) 	&=& -\log_2 \left[ \mathrm{Guess}(X|Y) \right] \\ \notag
&=& -\log_2 \left[ \sum_y p(y) \mathrm{Guess}(X|Y=y) \right] \\ \notag 
&=&	-\log_2 \left[ \sum_y p(y) \max_x \left\lbrace p(x|y) \right\rbrace \right] \ ,
\label{eqn:hmin}
\end{eqnarray}

\noindent
where $p(y)$ describes as above the probability for the outcome $y$ of the random variable $Y$ and $p(x|y)$ describes the conditional probability for the outcome $x$ of the random variable $X$, conditioned on the outcome $y$ of the random variable $Y$. 

The entropy can be understood as the amount of information, the random number is containing. Our assumption is, that all events in the past can be known by an external adversary. We will calculate the conditional entropy of the next produced output $Y$. Only the knowledge about the previous produced output $X$ is relevant.

In the section~\ref{disc} all coincidences will be discarded. That means the only outputs are the bits $\mathbf{1}$ and $\mathbf{0}$. An outcome produced by detector $\adet$ is labeled as $x=\adet$. In section~\ref{incl} a coincidence will also be treated as an output of the generator. A coincidence as output is labeled as $x=\adet\bdet$.\\

Before we start to calculate all the required probabilities in Eqn.~\ref{eqn:shan} and \ref{eqn:hmin} for the case of discarding and not discarding coincidences, one function has to be defined. Eqn.~\ref{eq:P_coinc}, which gives the the probability that detector $\bdet$ clicks within the time interval $[-\tau_{\mathrm{cw}},\tau_{\mathrm{cw}}]$, conditioned on a click at $t=0$ on detector $\adet$, can be generalized to any time interval $[t_1,t_2]$ with $t_1 ,t_2 \in \mathbb{R} \ \land \ t_2 \geq t_2 \ \land \ (t_2-t_1) \leq \tau_{\mathrm{dead}}$:

\begin{equation}
P_{\mathrm{click}}^\bdet([t_1,t_2]):= \lambda_{\mathrm{click}}^\bdet \int_{t_1}^{t_2} g^{(2)}(\tau) \ d\tau \ .
\label{eq:P_click}
\end{equation}

\noindent
The function $P_{\mathrm{click}}^\bdet([t_1,t_2])$ is the probability that detector $\bdet$ clicks within the interval $[t_1,t_2]$ when there is a click at $t=0$ on detector $\adet$.

\subsection{Conditional probability of consecutive bits (discarding coincidences)}
\label{disc}

Now let us calculate the functions $p(y)$ and $p(x|y)$ if we discard coincidences, that means the only outcome of the random variables $X$ and $Y$ are $\mathbf{1}$ and $\mathbf{0}$.\\
The probabilities $p(y)$ are very easy to calculate:

\begin{equation}
p(\adet)= 
\frac{\lambda_{\mathrm{bit}}^{\adet}}{\lambda_{\mathrm{bit}}^\adet + \lambda_{\mathrm{bit}}^\bdet} \qquad \text{with}\ \adet, \bdet \in \{R,T\} \ \land \ \adet \neq \bdet \ .
\end{equation}

\noindent
The calculation of the conditional probabilities $p(\bdet|\adet)$ and $p(\adet|\adet)$ is more complicated.\\
Before we do that we define $p_{\mathrm{out} \tau_{\mathrm{dead}}}^\adet$, i.e. the probability if both detectors are outside dead time, the next output is $x=\adet$. In principle the beam splitter ``decides'' where the next photon will occur, but it has to be considered, that the detectors have a quantum efficiency and that the next click might end in a coincidence:

\begin{subequations}
\begin{eqnarray}
\notag
p_{{\mathrm{out}}\tau_{\mathrm{dead}}}^\adet &=& \underbrace{ p_\adet \eta_{\mathrm{qe}} P_0^\bdet(\tau_{\mathrm{cw}}) }_{(1)} + 
\underbrace{		\left[
\underbrace{	 \left( 1-\eta_{\mathrm{qe}} \right) }_{(2)} + 
\underbrace{ \eta_{\mathrm{qe}}p_\adet \left( 1-P_0^\bdet(\tau_{\mathrm{cw}}) \right) }_{(3)} +
\underbrace{ \eta_{\mathrm{qe}}p_\bdet \left( 1-P_0^\adet(\tau_{\mathrm{cw}}) \right) }_{(4)} 
\right]		}_{=\left[1- \eta_{\mathrm{qe}} \left( p_\adet P_0^\bdet(\tau_{\mathrm{cw}})+p_\bdet P_0^\adet(\tau_{\mathrm{cw}}) \right) \right]}\cdot \\
&& \cdot p_\adet \eta_{\mathrm{qe}} P_0^\bdet(\tau_{\mathrm{cw}}) + \underbrace{ \dots }_{(5)} \\
\label{eq:p_out_taud_case1_b}
 &=& p_\adet \eta_{\mathrm{qe}} P_0^\bdet(\tau_{\mathrm{cw}})
\sum_{j=0}^\infty \left[ \underbrace{ 
1- \eta_{\mathrm{qe}} \left( p_\adet P_0^\bdet(\tau_{\mathrm{cw}})+p_\bdet P_0^\adet(\tau_{\mathrm{cw}}) \right) }_{< 1} \right]^j \\
\label{eq:p_out_taud_case1_c}
 &=& p_\adet \eta_{\mathrm{qe}} P_0^\bdet(\tau_{\mathrm{cw}}) \frac{1}{\eta_{\mathrm{qe}} \left( p_\adet P_0^\bdet(\tau_{\mathrm{cw}})+p_\bdet P_0^\adet(\tau_{\mathrm{cw}}) \right)}\\
\label{eq:p_out_taud_case1_d} 
 &=& \frac{p_\adet P_0^\bdet(\tau_{\mathrm{cw}})}{p_\adet P_0^\bdet(\tau_{\mathrm{cw}})+p_\bdet P_0^\adet(\tau_{\mathrm{cw}})} \ ,
\end{eqnarray}
\end{subequations}

\noindent
in which

\begin{table}[H]
\centering
\begin{tabular}{|c|l|} \hline
\textbf{Term} & \textbf{Probability, that ...}
\\ \hline\hline
\textbf{(1)}	& a photon leads to a click on detector $\adet$ and within $\tau_{\mathrm{cw}}$\\
& detector $\bdet$ will not click.\\
& $\Rightarrow$ the outcome $\adet$ is produced.		\\\hline
\textbf{(2)}	& the photon will not click either on detector $\adet$ or $\bdet$.\\
& $\Rightarrow$ no output is produced. \\\hline
\textbf{(3)}	& the photon clicks on detector $\adet$ and detector $\bdet$ clicks also \\
& within $\tau_{\mathrm{cw}}$.\\
& $\Rightarrow$ a coincidence occurs. \\\hline
\textbf{(4)}	& $(3)$ with permuted $\adet$, $\bdet$.\\
& $\Rightarrow$ a coincidence occurs.	\\\hline	
\textbf{(5)}	& $(m-1)$ photons did not produce an outcome (due to $(2)$,$(3)$,$(4)$)\\
& and the $m$th photon produce the outcome $\adet$ (with $m\in \mathbb{N} \backslash \{ 1 \}$).\\
& $\Rightarrow$ the outcome $\adet$ is produced. 	\\\hline			
\end{tabular}
\end{table}

To simplify the infinite sum in Eqn.~\ref{eq:p_out_taud_case1_b} we use the geometric series.\\
With the Eqn. \ref{eq:P_0(t1,t2)}, \ref{eq:P_coinc}, \ref{eq:P_click} and \ref{eq:p_out_taud_case1_d} the conditional probabilities can be calculated:

\begin{eqnarray}
\label{eq:p_detb_under_deta}
p(\bdet|\adet) &\approx & 	\underbrace{ \frac{ P_{\mathrm{click}}^\bdet([\tau_{\mathrm{cw}},\tau_{\mathrm{dead}}-\tau_{\mathrm{cw}}]) }
{1-P_{\mathrm{coinc}}^\bdet} }_{(1)} + \\ \notag
&&	\underbrace{ \left( 1-\frac{ P_{\mathrm{click}}^\bdet([\tau_{\mathrm{cw}},\tau_{\mathrm{dead}}]) }
{1-P_{\mathrm{coinc}}^\bdet} \right) }_{(2)} p_{\mathrm{out} \tau_{\mathrm{dead}}}^\bdet + \\ \notag				
&&	\underbrace{ \int_{\tau=0}^{\tau_{\mathrm{cw}}} \frac{ \lambda_{\mathrm{click}}^\bdet g^{(2)}(\tau_{\mathrm{dead}}-\tau_{\mathrm{cw}}+\tau) }
{1-P_{\mathrm{coinc}}^\bdet} P_0^\adet([\tau_{\mathrm{cw}}-\tau,\tau_{\mathrm{cw}}])\ d\tau }_{(3)} + \\ \notag
&&	\underbrace{ \int_{\tau=0}^{\tau_{\mathrm{cw}}} \frac{ \lambda_{\mathrm{click}}^\bdet g^{(2)}(\tau_{\mathrm{dead}}-\tau_{\mathrm{cw}}+\tau) }
{1-P_{\mathrm{coinc}}^\bdet} (1-P_0^\adet([\tau_{\mathrm{cw}}-\tau,\tau_{\mathrm{cw}}]))\ d\tau 
}_{(4)} \ p_{\mathrm{out} \tau_{\mathrm{dead}}}^\bdet \\
\label{eq:p_deta_under_deta}	
p(\adet|\adet)&\approx & \underbrace{ \left( 1- \frac{ P_{\mathrm{click}}^\bdet([\tau_{\mathrm{cw}},\tau_{\mathrm{dead}}]) }{1-P_{\mathrm{coinc}}^\bdet} \right)
p_{\mathrm{out} \tau_{\mathrm{dead}}}^\adet }_{(2)} + \\ \notag
&& 	\underbrace{ \int_{\tau=0}^{\tau_cw} \frac{ \lambda_{\mathrm{click}}^\bdet g^{(2)}(\tau_{\mathrm{dead}}-\tau_{\mathrm{cw}}+\tau)}
{1-P_{\mathrm{coinc}}^\bdet} ( 1- P_0^\adet([\tau_{\mathrm{cw}}-\tau,\tau_{\mathrm{cw}}]))
\ d\tau }_{(4)} \ p_{\mathrm{out} \tau_{\mathrm{dead}}}^\adet 
\end{eqnarray}

\begin{figure}[H]
\centering
\subfigure[Term (1)]{\includegraphics[width=0.7 \columnwidth ]{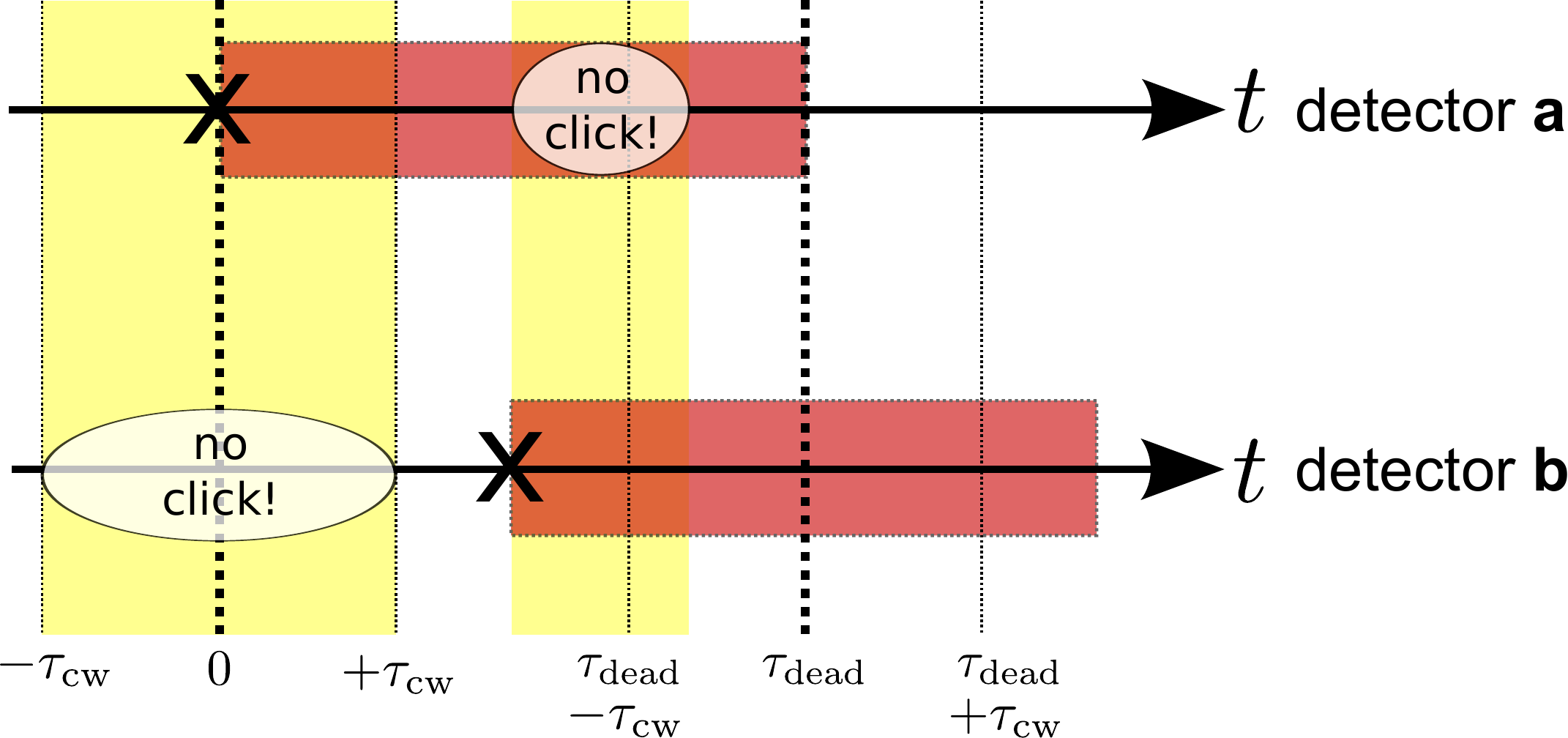}} \vfill
\subfigure[Term (2)]{\includegraphics[width=0.7 \columnwidth]{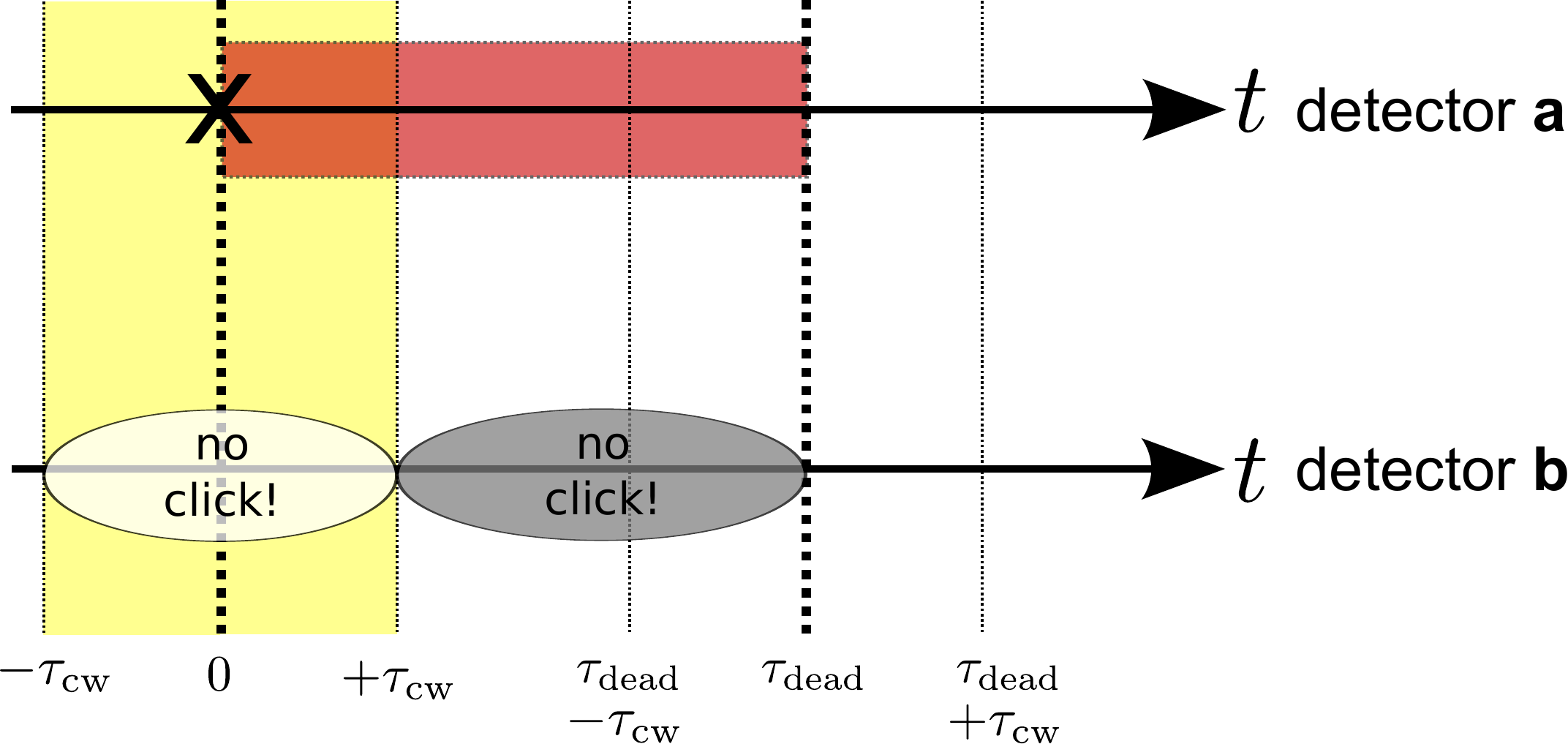}} \vfill
\subfigure[Term (3)]{\includegraphics[width=0.7 \columnwidth ]{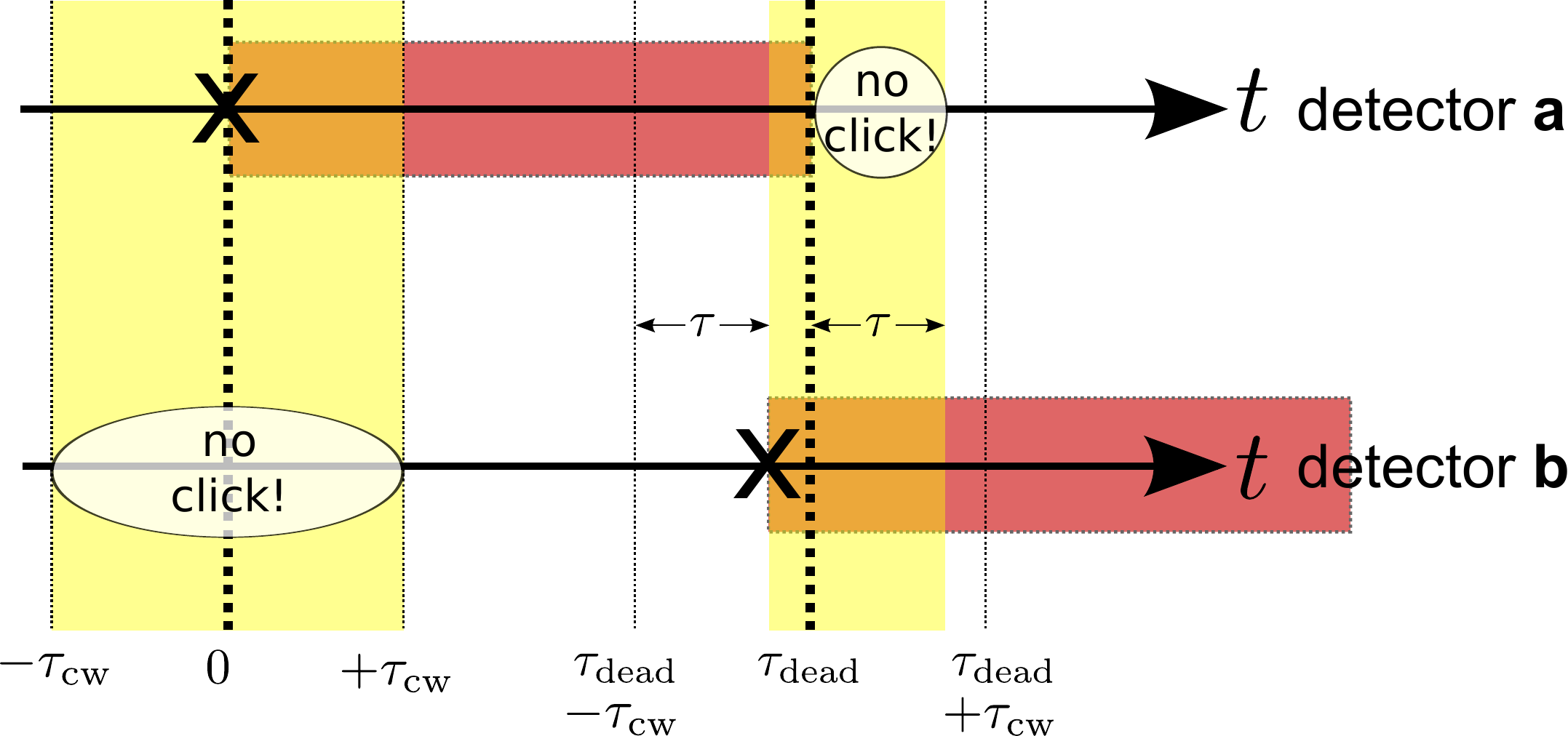}} \vfill
\subfigure[Term (4)]{\includegraphics[width=0.7 \columnwidth]{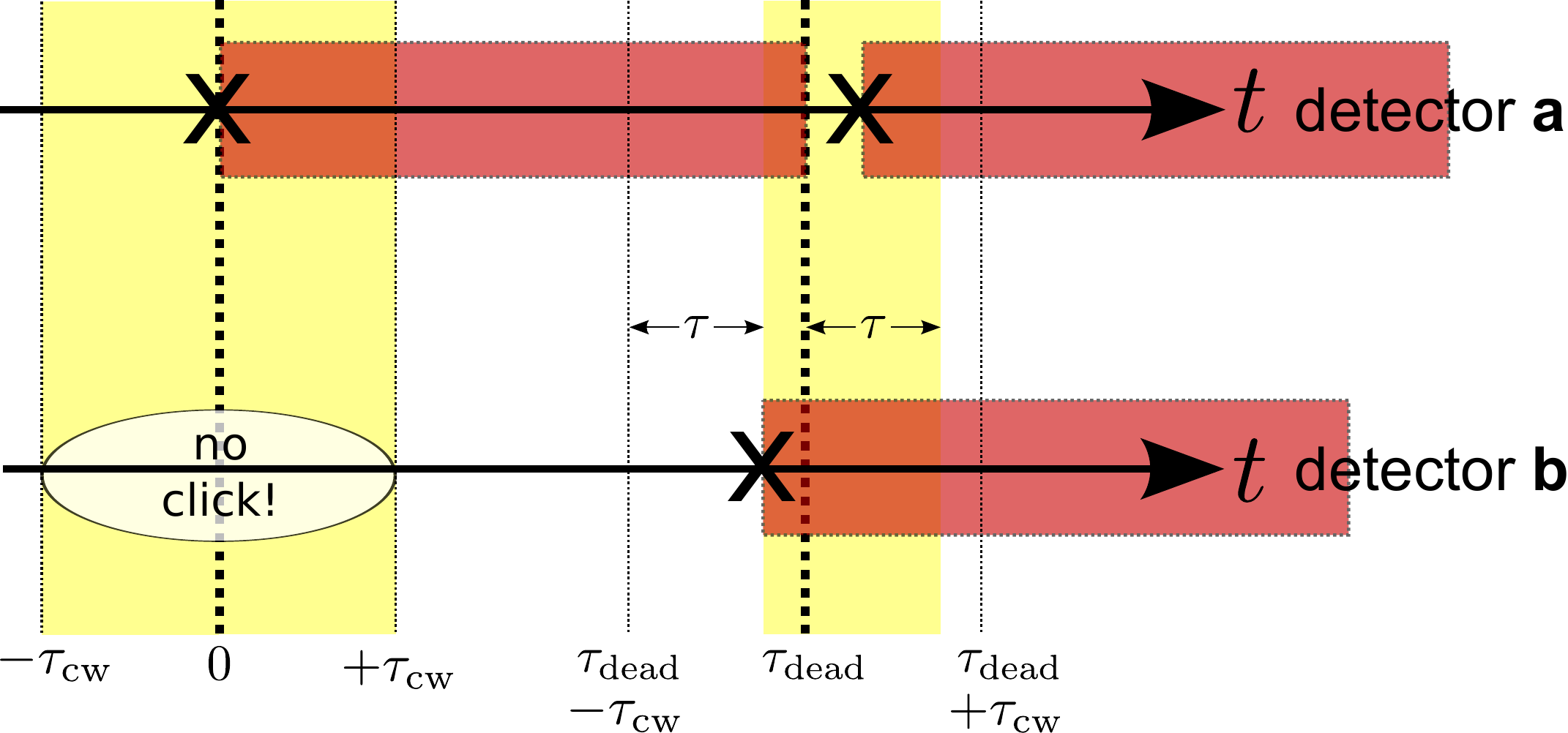}}
\caption{Visualization of four terms defined in Eqn.~\ref{eq:p_detb_under_deta}, \ref{eq:p_deta_under_deta}, \ref{eq:p_detb_under_deta_coinc}, \ref{eq:p_deta_under_deta_coinc} and \ref{eq:p_detadetb_under_deta_coinc} to calculate the conditional probabilities.}
\label{Fig:terms_in_cond_prob}
\end{figure}

\noindent
All these probabilities are conditioned to the fact that detector $\adet$ clicks at $t=0$ \emph{and} detector $\bdet$ does not click within the time interval $[-\tau_{\mathrm{cw}},\tau_{\mathrm{cw}}]$, since otherwise the outcome $x=\adet$ would not have been produced. The last-mentioned condition manifests in the calculations through the term $1/(1-P_{\mathrm{coinc}}^\bdet)$. The meaning the four occurring terms in Eqn.~\ref{eq:p_detb_under_deta} and \ref{eq:p_deta_under_deta} (and also Eqn.~\ref{eq:p_detb_under_deta_coinc}, \ref{eq:p_deta_under_deta_coinc} and \ref{eq:p_detadetb_under_deta_coinc}) are explained in Tab.~\ref{tab:terms_in_cond_prob} and are visualized in Fig.~\ref{Fig:terms_in_cond_prob}.

\begin{table}[H]
\centering
\begin{tabular}{|c||l|} \hline
\textbf{Term}		& \textbf{Conditional probability, that ...} \\ \hline\hline
\textbf{(1)}	& within the time interval $[\tau_{\mathrm{cw}},\tau_{\mathrm{dead}}-\tau_{\mathrm{cw}}]$ detector $\bdet$ clicks.\\
& $\Rightarrow$ definitely outcome $x=\bdet$ is produced. \\ \hline
\textbf{(2)}	& detector $\bdet$ clicks not within the interval $[\tau_{\mathrm{cw}},\tau_{\mathrm{dead}}]$.\\
& $\Rightarrow$ both detectors are outside dead time after $t=\tau_{\mathrm{dead}}$. \\ \hline
& detector $\bdet$ clicks at time $t=\tau_{\mathrm{dead}}-\tau_{\mathrm{cw}}+\tau$ (with $\tau \in [0,\tau_{\mathrm{cw}}]$) and within the interval \\
\textbf{(3)}	& $[\tau_{\mathrm{dead}},\tau_{\mathrm{dead}}+\tau]$ detector $\adet$ clicks not.\\
& $\Rightarrow$ definitely outcome $x=\bdet$ is produced. \\ \hline
& detector $\bdet$ clicks at time $t=\tau_{\mathrm{dead}}-\tau_{\mathrm{cw}}+\tau$ (with $\tau \in [0,\tau_{\mathrm{cw}}]$) and within the interval \\
\textbf{(4)}	& $[\tau_{\mathrm{dead}},\tau_{\mathrm{dead}}+\tau]$ detector $\adet$ clicks.\\
& $\Rightarrow$ a coincidence is occurring. \\ \hline	
\end{tabular}
\caption{Explanation of the four terms occurring in Eqn.~\ref{eq:p_detb_under_deta}, \ref{eq:p_deta_under_deta}, \ref{eq:p_detb_under_deta_coinc}, \ref{eq:p_deta_under_deta_coinc} and \ref{eq:p_detadetb_under_deta_coinc}.}
\label{tab:terms_in_cond_prob}
\end{table}

We introduce Eqn.~\ref{eq:p_detb_under_deta} and \ref{eq:p_deta_under_deta} -- here we introduce an approximation in the last line of each equation. The case described in Term~(4) leads to a coincidence. The detector clicks, which produces those coincidence, causes a dead time on each detector. But the detector clicks are not necessarily occurring at the same point in time, which means, that the ending of the dead time of both detectors is not happening at the same time. For simplicity it is assumed, that both detectors dead times end exactly at the same time after a coincidence. This is justified, since the dead time is very long against the introduced coincidence window.\\

The calculated transition probabilities for output $y=\adet$ followed by $x=\bdet$ ($p(\bdet|\adet)$) and for output $y=\adet$ followed $x=\adet$ ($p(\adet|\adet)$) are shown in Fig.~\ref{fig:entropy}a for both light sources.

For small light intensities the chances for the next output are approximately equal for bit $\mathbf{1}$ and $\mathbf{0}$. As the intensity is increased, it is more likely that the a $\mathbf{0}$ is followed by a $\mathbf{1}$ or vice versa, which makes it easier for an external adversary to guess the next outcome. In the case of the single photon source the discrepancy between the transition probabilities is lower than in the case of the laser, which results in an advantage of the single photon source. For infinitely high laser intensities the probability for a bit-flip goes to one. In this case the generators output would be: $\adet$,$\bdet$,$\adet$,$\bdet$,$\adet$,$\bdet$,$\adet$,$\bdet$,$\dots$ . 

\subsection{Conditional probability of consecutive outcomes (including coincidences)}
\label{incl}

If as well as the outcomes $x=\adet$, a coincidence ($x=\adet \bdet$) is treated also as a legitimately output, a few functions have to be prepared.\\
The first difference is the probability $p(y)$ of an arbitrary outcome:

\begin{equation}
p(y)= 
\frac{\lambda_{\mathrm{bit}}^{y}}{\lambda_{\mathrm{bit}}^\adet + \lambda_{\mathrm{bit}}^\bdet + \lambda_{\mathrm{bit}}^{\adet \bdet}}
\qquad \mathrm with \ y \in \{ \adet,\adet \bdet \} \ .
\end{equation}

\noindent
The functions $p_{\mathrm{out} \tau_{\mathrm{dead}}}^x$ (with $x \in \{\adet,\adet \bdet\}$), which denotes the probability to receive the outcome $x$ when both detectors are outside dead time, changes also:

\begin{eqnarray}
\label{eq:p_out_taud_case2_a}
p_{\mathrm{out} \tau_{\mathrm{dead}}}^\adet &=& \eta_{\mathrm{qe}} p_\adet P_0^\bdet(\tau_{\mathrm{cw}})\cdot 
\sum_{j=0}^\infty \left( \underbrace{1-\eta_{\mathrm{qe}}}_{<1} \right)^j = p_\adet P_0^\bdet(\tau_{\mathrm{cw}}) \\
\label{eq:p_out_taud_case2_ab}
p_{\mathrm{out} \tau_{\mathrm{dead}}}^{\adet \bdet} &=& \left[ \eta_{\mathrm{qe}} p_\adet ( 1-P_0^\bdet(\tau_{\mathrm{cw}}) ) +
\eta_{\mathrm{qe}} p_\bdet ( 1-P_0^\adet(\tau_{\mathrm{cw}}) ) \right]
\cdot \sum_{j=0}^\infty \left( \underbrace{1-\eta_{\mathrm{qe}}}_{<1}\right)^j \\ \notag
&=& 1 - p_\adet P_0^\bdet(\tau_{\mathrm{cw}}) - p_\bdet P_0^\adet(\tau_{\mathrm{cw}})
\end{eqnarray}

\noindent
And with the Eqn. \ref{eq:P_0(t1,t2)}, \ref{eq:P_coinc}, \ref{eq:P_click}, \ref{eq:p_out_taud_case2_a} and \ref{eq:p_out_taud_case2_ab} the conditional probabilities $p(x|y)$ denote in this case

\begin{eqnarray}
\label{eq:p_detb_under_deta_coinc}
p(\bdet|\adet) &=& \underbrace{ \frac{ P_{\mathrm{click}}^\bdet([\tau_{\mathrm{cw}},\tau_{\mathrm{dead}}-\tau_{\mathrm{cw}}]) }{1-P_{\mathrm{coinc}}^\bdet} }_{(1)} + \\ \notag
&&	\underbrace{ \left( 1- \frac{ P_{\mathrm{click}}^\bdet([\tau_{\mathrm{cw}},\tau_{\mathrm{dead}}]) }{1-P_{\mathrm{coinc}}^\bdet} \right) }_{(2)}
p_{\mathrm{out} \tau_{\mathrm{dead}}}^\bdet + \\ \notag				
&&	\underbrace{ \int_{\tau=0}^{\tau_{\mathrm{cw}}} \frac{ \lambda_{\mathrm{click}}^\bdet g^{(2)}(\tau_{\mathrm{dead}}-\tau_{\mathrm{cw}}+\tau) }
{1-P_{\mathrm{coinc}}^\bdet} P_0^\adet([\tau_{\mathrm{cw}}-\tau,\tau_{\mathrm{cw}}]) }_{(3)} \ d\tau
\end{eqnarray}

\begin{eqnarray}
\label{eq:p_deta_under_deta_coinc}								 		
p(\adet|\adet) &=& \underbrace{ \left( 1- \frac{ P_{\mathrm{click}}^\bdet([\tau_{\mathrm{cw}},\tau_{\mathrm{dead}}]) }{1-P_{\mathrm{coinc}}^\bdet} \right) }_{(2)}
p_{\mathrm{out} \tau_{\mathrm{dead}}}^\adet
\end{eqnarray}

\begin{eqnarray}
\label{eq:p_detadetb_under_deta_coinc}
p(\adet \bdet|\adet) &=& \underbrace{ \left( 1- \frac{ P_{\mathrm{click}}^\bdet([\tau_{\mathrm{cw}},\tau_{\mathrm{dead}}]) }{1-P_{\mathrm{coinc}}^\bdet} \right) }_{(2)} p_{\mathrm{out} \tau_{\mathrm{dead}}}^{\adet \bdet} + \\ \notag
&& \underbrace{ \int_{\tau=0}^{\tau_{\mathrm{cw}}} \frac{ \lambda_{\mathrm{click}}^\bdet g^{(2)}(\tau_{\mathrm{dead}}-\tau_{\mathrm{cw}}+\tau) }
{1-P_{\mathrm{coinc}}^\bdet} 
\left( 1-P_0^\adet([\tau_{\mathrm{cw}}-\tau,\tau_{\mathrm{cw}}]) \right) }_{(4)} \ d\tau
\end{eqnarray}

\begin{eqnarray}
\label{eq:p_deta_under_detadetb_coinc}
p(\adet|\adet \bdet) &\approx & p_{\mathrm{out} \tau_{\mathrm{dead}}}^\adet \\
\label{eq:p_detadetb_under_detadetb_coinc}
p(\adet \bdet|\adet \bdet) &\approx & p_{\mathrm{out} \tau_{\mathrm{dead}}}^{\adet \bdet}
 \ .
\end{eqnarray}

\noindent
The meaning of these four terms in Eqn.~\ref{eq:p_detb_under_deta_coinc}, \ref{eq:p_deta_under_deta_coinc} and \ref{eq:p_detadetb_under_deta_coinc} is described above in Tab.~\ref{tab:terms_in_cond_prob} and visualized in Fig.~\ref{Fig:terms_in_cond_prob}. Note, that in Eqn.~\ref{eq:p_deta_under_detadetb_coinc} and \ref{eq:p_detadetb_under_detadetb_coinc} is made the same approximation as above -- it is assumed, that the dead times, caused by the detector clicks of a coincidence, end at exactly the same time.\\

A plot of all transition probabilities $p(x|y)$ is given in Fig.~\ref{fig:entropy}b.

What we can see for the transitions $y=\adet$ to outcome $x$ (with $x\in\{\adet,\bdet,\adet\bdet\}$) is similar to the case without coincidences (Fig.~\ref{fig:entropy}a). On the bottom there is now also a small probability that the outcome $\adet$ is followed by a coincidence ($x=\adet\bdet$). The coincidences break the symmetry, which can be observed in the case without them. For the transitions $y=\adet\bdet$ to $x$ (with $x\in\{\adet,\bdet,\adet\bdet\}$) it can be seen at low incoming photon rates, that the output $x$ is with 50\% chance $x=\adet$ or $x=\bdet$. But as the incoming photon rate decreases the outputs $x=\adet,\bdet$ decreases and the output $x=\adet\bdet$ increases. When the incoming photon rate is $\lambda=6.73\cdot 10^8$~s$^{-1}$, all outputs $x$ become equal probable (1/3), which means that the occurrence of a coincidence makes it hard for an external adversary to guess the next output. In the limit case of infinitely high photon rate the chance of a coincidence goes to 100\%. So in the limit case the generator output consists of two different patterns: if we start with the output

\begin{itemize}
\item $\adet \rightarrow$ $\adet$,$\bdet$,$\adet$,$\bdet$,$\adet$,$\bdet$, \dots .
\item $\adet\bdet \rightarrow$ $\adet\bdet$,$\adet\bdet$,$\adet\bdet$,$\adet\bdet$,$\adet\bdet$,$\adet\bdet$, \dots .
\end{itemize}

\subsection{Shannon- and min-entropy with and without coincidences}

Now we have all tools to calculate the Shannon- and min-entropy per outcome (Eqn.~\ref{eqn:shan} and \ref{eqn:hmin}). These entropies are calculated for the case of discarding coincidences and not discarding coincidences for both light sources. In Fig.~\ref{fig:entropy}c shows the entropies per bit and also the entropy difference of the cases without and with discarding coincidences.

It can be obtained, that for very low incoming photon rates all entropies are nearly unity and for infinitely high photon rates the entropy functions goes to zero (according to the occurrence of patterns). The entropy per outcome is greater if coincidences are not discarded and also treated as an output. Interestingly, the Shannon entropy calculated for the laser reaches values greater than one (dashed green line). This is caused by the coincidence as third output. The entropy loss $\Delta \mathcal{H}$ between the cases discarding and not discarding coincidences reaches its maximum at an incoming photon rate $\lambda \approx 6\cdot 10^8$~s$^{-1}$. This can be understood by a look back to Fig.~\ref{fig:entropy}b, where at this photon rate all transition probabilities $\adet\bdet \rightarrow x$ become equal. But the important functions are the min-entropies per outcome, which give an lower entropy bound. If coincidences are discarded, the single photon source (solid blue line) has always a greater min-entropy than the laser (dashed blue line). With coincidences the single photon source (solid yellow line) is superior for incoming photon rates $0 < \lambda < 5.6 \cdot 10^7$~s$^{-1}$ and for photon rates $5.6 \cdot 10^7 < \lambda < \mathit{\Gamma}=10^8 $~s$^{-1}$ the laser (dashed yellow line) is superior.\\

The entropy \emph{rate} can be calculated through the entropy per outcome multiplied with the outcome rate. The latter is in the case of discarding coincidences $\lambda_{\mathrm{bit}}^\adet + \lambda_{\mathrm{bit}}^\bdet$ and in the case when coincidences are not discarded $\lambda_{\mathrm{bit}}^\adet + \lambda_{\mathrm{bit}}^\bdet + \lambda_{\mathrm{bit}}^{\adet\bdet}$. In Fig.~\ref{fig:entropy}d are shown the entropy rates and also the entropy rate difference of the cases not discarding and discarding coincidences.

For small photon rates the entropy rate is small (since the outcome rate is small). As the photon rate increases, the entropy rate increases, too. After the entropy rate reaches it maximum, it vanishes for infinitely high photon rates (since the entropy per outcome vanishes). The entropy rate with coincidences is for all photon rates higher than without coincidences and the entropy rate of the single photon source is for (almost) all photon rates higher than the laser ones. The loss of entropy rate by discarding coincidences is shown at the bottom of the plots. Again its maximum lies around $\lambda=6\cdot10^8$~s$^{-1}$ (cf. Fig.~\ref{fig:entropy}b). The inlay shows a zoom to the maxima of the min-entropy rates. In order to produce random digits fast, the generator should be driven at photon rates corresponding to these optimum points. The min-entropy rate maximum of the single photon source with coincidences (yellow solid line) denotes $\mathcal{H}_{\mathrm min}=9.23 \times 10^6$~s$^{-1}$ at a photon rate $\lambda=5.03\times 10^7$~s$^{-1} \mathrel{\widehat{=}} k=1.01\times 10^8$~$s^{-1}$ and without coincidences (blue solid line) $\mathcal{H}_{\mathrm min}=9.12 \times 10^6$~s$^{-1}$ at $\lambda=5.21\times 10^7$~s$^{-1} \mathrel{\widehat{=}} k=1.09\times 10^8$~s$^{-1}$. Note, that the optimum point of the single photon source with coincidences ($\lambda=5.03\times 10^7$~s$^{-1}$) lies still in the regime, where also the the min-entropy per outcome of the single photon source is superior against the laser (cf. Fig.~\ref{fig:entropy}c). The min-entropy rate maximum of the laser with coincidences (yellow dashed line) denotes $\mathcal{H}_{\mathrm min}=8.74 \times 10^6$~s$^{-1}$ at a photon rate $\lambda=6.14\times 10^7$~s$^{-1} \mathrel{\widehat{=}} k=1.60\cdot 10^8$~s$^{-1}$ and without coincidences (blue dashed line) $\mathcal{H}_{\mathrm min}=8.09 \times 10^6$~s$^{-1}$ at $\lambda=5.29\times 10^7$~s$^{-1} \mathrel{\widehat{=}} k=1.12\times 10^8$~s$^{-1}$.
If the four min-entropy optimum points of the single photon source and the laser are compared, it turns out that:

\begin{itemize}
\item without coincidences the entropy rate of the single photon source is $12.7\%$ greater than the rate of the laser.
\item with coincidences the entropy rate of the single photon source is $5.6\%$ greater than the rate of the laser.
\item with the laser the entropy rate would be $8.0\%$ greater if coincidences are not discarded.
\item with the single photon source the entropy rate would be $1.3\%$ greater if coincidences are not discarded.
\end{itemize}

As a last step, the entropy should be extracted from the raw bits. This is usually performed in a two-universal hashing procedure~\cite{tomamichel_itoit_2011}. 

Of course, the technical subtleties (the jitter and the dead-time) can be suppressed, when more detetors are introduced. Here, the QRNG could be improved by using $\sum_{j=0}^n 2^j$ beam splitters. That means in the output path of a beam splitter another beam splitter is placed. A generator based on this principle could produce by one single incoming photon $n+1$ bits. This scheme was realized in integrated optics~\cite{graefe_nphot_2014}. Very likely this configuration makes a laser superior to a single photon source. Furthermore, this increases the costs of the generator also by $\sum_{j=0}^n 2^j$.

Another option is to supply a pulsed light source. This configuration is well analyzed for the case of a laser source~\cite{frauchiger_a_2013}.

\begin{appendix}

\section{Relevant input parameters for the calculations in the main manuscript}

The input parameters for this study are presented in Table~\ref{table:parameter}.

\begin{table}
\caption{Relevant input parameters for the calculations in the main manuscript}
\begin{tabular}{|l|l|r|}
\hline
Name of parameter & Symbol & Value \\ \hline
Quantum efficiency & $\eta$ & 60\% \\ \hline
Incident photon flux & $\lambda$ & 10$^{5}$-10$^{12}$~s$^{-1}$ \\ \hline
Dead time & $\tau_{\mathrm dead}$ & 50~ns \\ \hline
Beam splitter ratio& $p_\adet$ & 0.5 (`fair' beam splitter) \\ \hline
Pump rate (single emitter) & $k$ & 10$^{1}$-10$^{12}$~per second \\ \hline
Life time (single emitter) & $T_1$ & 10~ns $\Leftrightarrow$ $\mathit{\Gamma}_1$=10$^8$~s$^-1$\\ \hline
\end{tabular}
\label{table:parameter}
\end{table}

\label{sect:Lm_Pm}

\section{Calculation of $P_{\mathrm m}$ and $L_{\mathrm m}$}

In the following we present the source code to calculate the $P_{\mathrm m}$ and $L_{\mathrm m}$ in \texttt{Mathematica} 9.0.1:

\noindent\(\pmb{\text{dir}=\text{{``}$\sim $/science/projects/trng/mathematica/calculated$\_$functions/{''}};}\\
\pmb{\text{(* define directory to save files there *)}}\)

\subsection*{Definition of \(g^{(2)}\)($\tau $)- and \(G^{(2)}\)($\tau $)- functions of laser and single photon source :}

\noindent\(\pmb{\text{g2L}[\tau \_]\text{:=}1 ;}\\
\pmb{\text{(*}\ \text{define}\ \text{g2}-\text{function}\ \text{of}\ \text{laser}\ \text{*)}}\)

\noindent\(\pmb{\text{G2L}[\tau \_]\text{:=}\text{g2L}[\tau ](\eta \lambda )^2;}\\
\pmb{\text{(*}\ \text{define}\ \text{G2}-\text{function}\ \text{of}\ \text{laser}\ \text{*)}}\)

\noindent\(\pmb{k[\lambda \_]\text{:=}\frac{\lambda \Gamma }{\Gamma -\lambda };}\\
\pmb{\text{(* express the pumping rate k as a function of emitted photon rate $\lambda $ *)}}\)

\noindent\(\pmb{\text{g2SPS}[\tau \_]\text{:=}1-\text{Exp}[-(k[\lambda ]+\Gamma )\tau ];}\\
\pmb{\text{(*}\ \text{g2}-\text{function}\ \text{of}\ \text{single}\ \text{photon}\ \text{source}\ \text{*)}}\)

\noindent\(\pmb{\text{G2SPS}[\tau \_]\text{:=}\text{g2SPS}[\tau ] (\eta \lambda )^2;}\\
\pmb{\text{(*}\ \text{G2}-\text{function}\ \text{of}\ \text{single}\ \text{photon}\ \text{source}\ \text{*)}}\)

\subsection*{Select here: laser or single photon source}

\noindent\(\pmb{\text{Laser} = \text{False};}\\
\pmb{\text{(*}\ \text{Select}\ \text{here}:\ \text{laser}\ (\text{Laser}=\text{True})\ \text{or}\ \text{single}\ \text{photon}\ \text{source}\ (\text{Laser}=\text{False})\ \text{*)}}\)

\noindent\(\pmb{\text{If}[\text{Laser},}\\
\pmb{\text{ }\text{g2}[\tau \_] = \text{g2L}[\tau ]; \text{G2}[\tau \_] = \text{G2L}[\tau ]; \text{name}=\text{{``}laser{''}};,}\\
\pmb{\text{ }\text{g2}[\tau \_] = \text{g2SPS}[\tau ]; \text{G2}[\tau \_] = \text{G2SPS}[\tau ]; \text{name}=\text{{``}sps{''}};];}\)

\subsection*{JK-formalism}

\noindent\(\pmb{J[\tau \_]=\frac{\text{G2}[\tau ]}{\eta \lambda };}\\
\pmb{\text{(*}\ \text{define}\ \text{the}\ \text{photon}\ \text{number}\ \text{density}\ J(\tau)\ \text{*)}}\)

\noindent\(\pmb{\text{Jtilde}[\text{p$\_$}]= \text{LaplaceTransform}[J[\tau ],\tau ,p];}\\
\pmb{\text{(*}\ \text{Laplace}-\text{transformation}\ \text{of}\ J(\tau)\ \text{to}\ p-\text{space}\ \text{*)} }\)

\noindent\(\pmb{\text{Ktilde}[\text{p$\_$}]=\frac{\text{Jtilde}[p]}{1+\text{Jtilde}[p]};}\\
\pmb{\text{(*}\ \text{calculate}\ \text{the}\ \text{function}\ K(p)\ \text{in}\ p-\text{space}\ \text{*)}}\)

\noindent\(\pmb{K[\tau \_]= \text{FullSimplify}[\text{InverseLaplaceTransform}[\text{Ktilde}[p],p,\tau ],}\\
\pmb{\{\lambda >0,\Gamma >0,\tau \geq 0,\eta >0,\lambda \in \text{Reals},\Gamma \in \text{Reals},\tau \in \text{Reals}, \eta \in \text{Reals}\}];}\\
\pmb{\text{(*}\ \text{Inverse}\ \text{Laplace}-\text{transformation}\ \text{of}\ K(p)\ \text{to}\ \text{revieve}\ K(\tau)\ \text{in}\ \text{time}-\text{space}\
\text{*)}}\)

\noindent\(\pmb{L[1,\tau \_]=K[\tau ];}\\
\pmb{\text{(* this is only nomenclature *)}}\)

\subsection*{Calculate the functions \(L_m(\tau )\) and export them to files}

\noindent\(\pmb{L[1,\tau ]>>\text{dir} <>\text{{``}Lm/L$\_${''}}<>\text{{``}1{''}}<>\text{{``}.nb{''}}}\\
\pmb{\text{(*}\ \text{export}\ L_1(\tau)\ \text{to}\ \text{file}\ \text{*)}}\)

\noindent\(\pmb{\text{Dynamic}[m+1]}\\
\pmb{\text{(* shows which function is calculated right now *)}}\)

\noindent\(\pmb{\text{Do}[L[m+1,\tau \_]=\text{Simplify}[\text{Integrate}[L[m,\tau -\tau \tau ] L[1,\tau \tau ],\{\tau \tau ,0,\tau \}],}\\
\pmb{\{\lambda >0,\lambda \leq \Gamma ,\Gamma >0,\tau \geq 0,\eta >0,\eta \leq 1,\lambda \in \text{Reals},\Gamma \in \text{Reals},\tau \in 
\text{Reals}, \eta \in \text{Reals}\}];}\\
\pmb{L[m+1,\tau ]>>\text{dir} <>\text{{``}Lm/L$\_${''}}<>\text{ToString}[m+1]<>\text{{``}.nb{''}}}\\
\pmb{,\{m,1,14\}]}\\
\pmb{\text{(*}\ \text{Calculate}\ \text{and}\ \text{export}\ \text{iteratively}\ \text{the}\ \text{functions}\ L_{m+1}(\tau)\ \text{*)}}\)

\subsection*{Calculate the functions \(P_m(T)\) and export them to files}

\noindent\(\pmb{P[0,\text{T$\_$}]=\text{FullSimplify}[1-\text{Integrate}[(L[1,\tau ]),\{\tau ,0,T\}],}\\
\pmb{\{\lambda >0,\Gamma >0,T\geq 0,\eta >0,\lambda \in \text{Reals},\Gamma \in \text{Reals},T \in \text{Reals}, \eta \in \text{Reals}\}];}\\
\pmb{P[0,T]>>\text{dir} <>\text{{``}Pm/P$\_${''}}<>\text{{``}0{''}}<>\text{{``}.nb{''}}}\\
\pmb{\left.\text{(*}\ \text{Calculate}\ \text{and}\ \text{export}\ \text{the}\ \text{function}\ P_0(\tau)\ \text{*)}\right)}\)

\noindent\(\pmb{\text{Dynamic}[m]}\\
\pmb{\text{(* shows which function is calculated right now *)}}\)

\noindent\(\pmb{\text{Do}[P[m,\text{T$\_$}]=\text{Simplify}[\text{Integrate}[L[m,\tau \tau ] (1-\text{Integrate}[(L[1,\tau ]),\{\tau ,0,T-\tau \tau
\}]),\{\tau \tau ,0,T\}],}\\
\pmb{\{\lambda >0,\Gamma >0,T\geq 0,\eta >0,\lambda \in \text{Reals},\Gamma \in \text{Reals},T \in \text{Reals}, \eta \in \text{Reals}\}];}\\
\pmb{P[m,T]>>\text{dir} <>\text{{``}Pm/P$\_${''}}<>\text{ToString}[m]<>\text{{``}.nb{''}}}\\
\pmb{,\{m,1,11\}]}\\
\pmb{\left.\text{(*}\ \text{Calculate}\ \text{and}\ \text{export}\ \text{iteratively}\ \text{the}\ \text{functions}\ P_m(\tau)\ \text{*)}\right)}\)

\section{Measuring the temporal jitter of the single photon detector}

\begin{figure}[H]
\begin{center}
\includegraphics[width=0.9 \columnwidth]{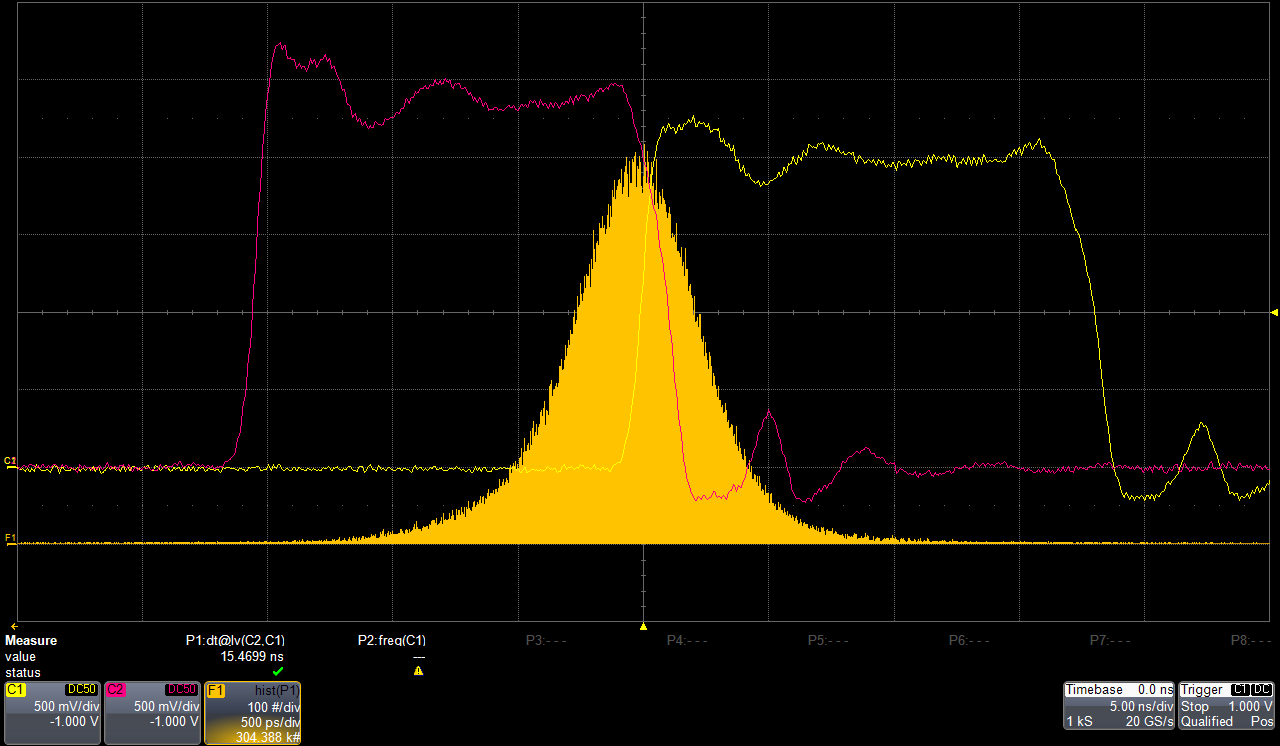}
\caption{Electrical jitter of the utilized single photon counting modules.}
\label{fig:jitter}
\end{center}
\end{figure}

\end{appendix}

\end{document}